\documentclass[aps,prr,twocolumn,amsmath,amssymb,superscriptaddress,longbibliography]{revtex4-2}
\usepackage[english]{babel}
\usepackage{xcolor}
\usepackage{amssymb} 
\usepackage{dcolumn}
\usepackage{bm}
\usepackage{graphicx}
\usepackage{caption}     
\usepackage{amsmath}
\usepackage{braket}
\usepackage{ragged2e}
\usepackage{graphicx}        

\graphicspath{{pict/}{}}

\usepackage[normalem]{ulem}

\usepackage{dcolumn}
\usepackage{bm}
\usepackage[pdfstartview=FitH, CJKbookmarks=true, bookmarksnumbered=true, bookmarksopen=true, colorlinks=true, pdfborder=001, citecolor=blue, linkcolor=blue, urlcolor=blue, linktocpage=true] {hyperref}

\begin{document}

\title{Interaction-enhanced photon blockade and Bell-state protection in microwave-shielded polar molecules}

\author{Yun Chen}
\thanks{These authors contributed equally to this work.}
\affiliation{Guangdong Provincial Key Laboratory of Quantum Metrology and Sensing $\&$ School of Physics and Astronomy, Sun Yat-Sen University (Zhuhai Campus), Zhuhai 519082, China}

\author{Geng Zhao}
\thanks{These authors contributed equally to this work.}
\affiliation{Guangdong Provincial Key Laboratory of Quantum Metrology and Sensing $\&$ School of Physics and Astronomy, Sun Yat-Sen University (Zhuhai Campus), Zhuhai 519082, China}

\author{Huanhuan Wei}
\affiliation{College of Sciences/State Key Laboratory of Advanced Energy Storage Materials and Technology, Shihezi University, Xinjiang, Shihezi, 832003, China.}

\author{Jingjun You}
\affiliation{Guangdong Provincial Key Laboratory of Quantum Metrology and Sensing $\&$ School of Physics and Astronomy, Sun Yat-Sen University (Zhuhai Campus), Zhuhai 519082, China}

\author{ Haoran Jia}
\affiliation{Guangdong Provincial Key Laboratory of Quantum Metrology and Sensing $\&$ School of Physics and Astronomy, Sun Yat-Sen University (Zhuhai Campus), Zhuhai 519082, China}

\author{Jing Tang}
\email{jingtang@gdut.edu.cn}
\affiliation{School of Physics and Optoelectronic Engineering, Guangdong University of Technology, Guangzhou 510006, China}
\affiliation{Guangdong Provincial Key Laboratory of Sensing Physics and System Integration Applications, Guangdong University of Technology, Guangzhou, 510006, China}

\author{Su Yi}
\email{yisu@nbu.edu.cn}
\affiliation{Institute of Fundamental Physics and Quantum Technology $\&$ School of Physical Science and Technology, Ningbo University, Ningbo, 315211, China}

\author{Yuangang Deng}
\email{dengyg3@mail.sysu.edu.cn}
\affiliation{Guangdong Provincial Key Laboratory of Quantum Metrology and Sensing $\&$ School of Physics and Astronomy, Sun Yat-Sen University (Zhuhai Campus), Zhuhai 519082, China}

\date{\today}

\begin{abstract}
Microwave shielding has recently emerged as a powerful tool for engineering interactions in ultracold polar molecules, yet its potential for controlling cavity quantum electrodynamics remains largely unexplored. Here, we investigate a molecular cavity quantum electrodynamics platform in which two microwave-shielded polar molecules are coupled to a single optical cavity mode and demonstrate that shielding-induced interactions provide a unified mechanism for both photon blockade and Bell-state protection. The anisotropic interaction reshapes the few-excitation spectrum by enhancing its anharmonicity, thereby suppressing multiphoton transitions and improving the single-photon purity by more than three orders of magnitude. The enhanced blockade is accompanied by the emergence of negative longitudinal spin correlations, revealing the interaction-induced suppression of simultaneous molecular excitations. We further show that the photon statistics are highly sensitive to the relative molecular configuration, with positional variations on the scale of the relative zero-point fluctuation substantially modifying the blockade performance. Beyond few-photon nonlinear optics, the same interaction protects an initially prepared molecular Bell state by dispersively decoupling molecular excitations from the lossy cavity mode, thereby suppressing cavity-mediated dissipation and slowing the fidelity decay. Our results establish microwave-shielded interactions as a unified interaction resource for engineering few-photon nonlinearities and protecting quantum states in molecular cavity-QED systems.
\end{abstract}

\maketitle 
\section{Introduction}

Ultracold polar molecules, with their tunable electric dipole moments and rich internal structures, provide a versatile platform for exploring exotic quantum phases~\cite{Carr_2009,Lahaye_2009,PhysRevLett.105.220406,PhysRevLett.109.266804,PhysRevLett.104.125302,PhysRevLett.98.060403,PhysRevLett.107.115301,PhysRevLett.103.155302,PhysRevLett.104.125301,PhysRevLett.98.060404}, quantum information processing~\cite{PhysRevLett.88.067901,C8SC02355G,Gregory2021}, many-body spin dynamics~\cite{Yan2013,Micheli2006,PhysRevLett.96.190401,PhysRevLett.110.075301,Yao2018,doi:10.1126/science.aam6299,Li2023,doi:10.1126/science.adf8999,Christakis2023}, and ultracold chemistry~\cite{Quemener2012,doi:10.1126/science.aam6299,Zhang2023}. A central challenge for molecular quantum gases is the suppression of short-range inelastic collisions, which has stimulated the development of external-field control, optical-lattice confinement~\cite{doi:10.1126/science.aau7230,Valtolina2020}, and microwave shielding~\cite{PhysRevLett.121.163401,PhysRevLett.121.163402}. In particular, microwave shielding employs microwave-dressed rotational states to generate repulsive interaction barriers and has been experimentally demonstrated to strongly suppress collisional losses~\cite{doi:10.1126/science.abg9502,PhysRevX.13.031032}, enabling evaporative cooling of fermionic molecules~\cite{Schindewolf2022} and contributing to the realization of molecular Bose-Einstein condensates~\cite{Bigagli2024}. Beyond collisional stabilization, microwave shielding offers a powerful means of engineering effective dipolar interactions, motivating theoretical studies of dipolar droplets, synthetic gauge fields, and other strongly interacting many-body phenomena~\cite{b8y9-yvz9,9cxl-d9zg,Deng2025,PhysRevLett.134.053001,k987-fqqw}. Despite these advances, the potential of microwave-shielded interactions for controlling light-matter interactions in cavity quantum electrodynamics remains largely unexplored. Exploiting microwave shielding as an interaction resource in cavity-coupled molecular systems could therefore open new opportunities for engineering quantum nonlinear optics and dissipative quantum dynamics.

\begin{figure}
 \includegraphics[width=0.98\columnwidth]{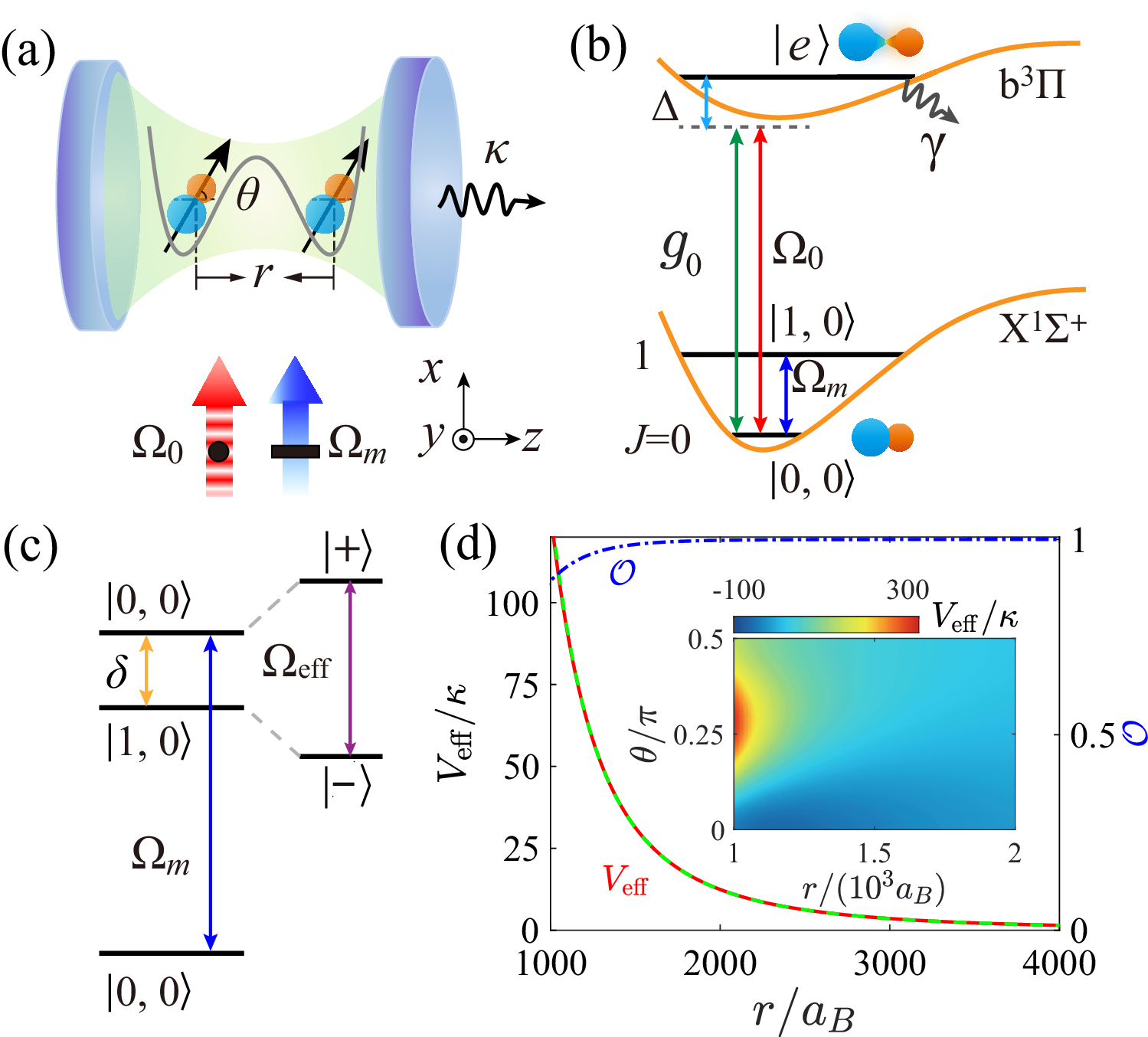}\captionsetup {labelformat=simple, labelsep=period, name=FIG.}  \caption{\label{model}  \justifying
(a) Sketch of the microwave-shielded cavity-coupled pair of polar molecules. (b) Molecular potentials and relevant electronic structure for engineering a rotational-state-resolved single-photon transition. (c) Schematic for creating $\pi$-polarized microwave-dressed molecules. (d) Effective potential $V_{\rm eff}$ from numerical diagonalization (solid line) and analytical expression (dashed line), together with state overlap ${\cal O}$ (dashed-dotted line) for $\theta=\pi/2$. The inset maps $V_{\rm eff}$ in the $r$-$\theta$ parameter plane for $\delta_r=0.1$ and $\Omega_m=2\pi\times5$ MHz.}
\end{figure}

High-purity single-photon sources are a central resource for quantum communication, distributed quantum computation, and quantum networks~\cite{RevModPhys.68.733,bennett2000quantum,O'Brien2009,Couteau2023,Couteau2023_2}. Photon blockade~\cite{PhysRevLett.79.1467} provides a fundamental mechanism for generating antibunched single-photon emission, in which the presence of one photon suppresses the resonant excitation of subsequent photons, leading to antibunched emission. Such nonclassical behavior generally originates from spectral anharmonicity~\cite{PhysRevA.49.R20,PhysRevLett.109.193602} or quantum interference between competing excitation pathways~\cite{PhysRevLett.104.183601,PhysRevA.83.021802,PhysRevLett.108.183601}. Photon blockade has been extensively investigated in atomic cavity-QED systems~\cite{Birnbaum2005}, cavity-coupled quantum dots~\cite{Faraon2008,PhysRevLett.114.233601,Reinhard2012}, and superconducting circuits~\cite{PhysRevLett.103.150503,PhysRevLett.106.243601,PhysRevLett.107.053602}. In contrast to these platforms, polar molecules provide highly tunable intermolecular interactions that can be engineered independently of the cavity field. Despite this advantage, exploiting engineered molecular interactions as the microscopic origin of photon blockade in cavity-QED systems remains largely unexplored. Microwave shielding provides precisely such an opportunity by enabling interaction-induced spectral anharmonicity without relying on intrinsic cavity nonlinearities or ultrastrong light-matter coupling.

Beyond nonclassical light generation, an equally important question is whether engineered molecular interactions can be exploited to control dissipative dynamics in cavity-QED systems. Recent experimental advances have demonstrated molecular qubits and entangled states in ultracold polar molecules~\cite{Gregory2021,doi:10.1126/science.adf8999,Ruttley2025}, motivating the development of strategies for preserving quantum coherence in open molecular platforms. In cavity QED, however, molecular excitations inevitably hybridize with lossy cavity photons, giving rise to cavity-mediated decoherence and the degradation of quantum correlations. Considerable efforts have therefore been devoted to protecting quantum states through decoherence-free subspaces~\cite{PhysRevLett.81.2594}, topological protection~\cite{RevModPhys.80.1083}, and reservoir engineering~\cite{Diehl2008,Verstraete2009,Reiter2012}. In contrast, microwave-shielded interactions provide a distinct interaction-based mechanism by modifying the energetic separation between molecular and photonic excitations, thereby suppressing cavity-mediated dissipation through dispersive decoupling. Although microwave shielding has been widely employed to suppress collisional losses and engineer interactions in ultracold molecular gases~\cite{doi:10.1126/science.abg9502,PhysRevX.13.031032,Schindewolf2022,Bigagli2024,PhysRevLett.130.183001}, its potential as a unified interaction resource for simultaneously engineering quantum nonlinear optics and protecting quantum states in molecular cavity-QED systems remains largely unexplored.

In this work, we show that microwave-shielded interactions provide a unified mechanism for engineering both photon blockade and quantum-state protection in molecular cavity QED. We consider a cavity-QED platform consisting of two microwave-shielded polar molecules coupled to a single optical cavity mode, where the anisotropic shielding interaction is controlled through the molecular geometry. We demonstrate that the interaction induces strong spectral anharmonicity in the few-excitation spectrum, thereby suppressing multiphoton transitions and enabling interaction-enhanced photon blockade. We further reveal how the interplay between intermolecular interactions, photon statistics, and spin correlations governs the microscopic origin of the blockade, as well as its sensitivity to the relative molecular configuration. Beyond few-photon nonlinear optics, we show that the same interaction suppresses cavity-mediated decoherence by dispersively decoupling molecular excitations from the cavity mode, thereby protecting an initially prepared Bell state. Our results establish microwave-shielded interactions as a unified interaction mechanism for engineering quantum nonlinear optics and dissipative quantum-state control in molecular cavity-QED systems.

\section{Model and Hamiltonian}

We consider a cavity-QED system consisting of two bialkali polar molecules confined at a controllable separation $r$ inside a high-finesse optical cavity with wavelength $\lambda=1145$ nm and decay rate $\kappa=2\pi\times40$ kHz~\cite{941q-5sdq,PhysRevLett.118.263601,kongkhambut2022observation}, as illustrated in Fig.~\ref{model}(a). A bias magnetic field applied along the $z$ axis defines the quantization axis for the rotational states labeled by the quantum numbers $(J,m_J)$. The molecular level structure is shown in Fig.~\ref{model}(b). We consider the optical transition from the rovibrational ground state
$|0,0\rangle\equiv X^1\Sigma^+|\upsilon=0,J=0,m_J=0\rangle$ to a long-lived excited bound state $|e\rangle\equiv b^3\Pi|\upsilon=0,J=1,m_J=1\rangle$, where $\upsilon$ denotes the vibrational quantum number~\cite{Gregory2024}. This transition is coupled to a $\sigma$-polarized cavity mode with single-molecule coupling strength $g_0$ and is weakly driven by a $\sigma$-polarized pump field with Rabi frequency $\Omega_0$ and detuning $\Delta$. Although the transition is electric-dipole forbidden in the absence of state mixing, it becomes weakly allowed through singlet-triplet mixing between the $b^3\Pi$ state and nearby $A^1\Sigma^+$ potentials~\cite{Gregory2024}, resulting in a narrow natural linewidth $\gamma$ that is well suited for cavity-QED experiments.

Direct dipole-dipole interactions (DDIs) involving the optical states $|0,0\rangle$ and $|e\rangle$ are generally too weak to generate appreciable optical nonlinearities. Instead, the essential nonlinear resource in our scheme originates from microwave shielding, which transforms the rotational ground-state manifold into a strongly interacting microwave-dressed channel featuring an anisotropic van der Waals-like repulsive core~\cite{PhysRevLett.121.163401,PhysRevLett.121.163402,doi:10.1126/science.abg9502,chen2023field,PhysRevLett.130.183001}. Specifically, the molecules are illuminated by a $\pi$-polarized microwave field propagating perpendicular to the cavity axis, which couples the rotational states $|0,0\rangle \leftrightarrow |1,0\rangle$ with Rabi frequency $\Omega_m$ and detuning $\delta$, as illustrated in Fig.~\ref{model}(c). Here, $|1,0\rangle\equiv X^1\Sigma^+|\upsilon=0,J=1,m_J=0\rangle$. The $m_J=\pm1$ sublevels remain uncoupled by the $\pi$-polarized microwave field, while higher rotational transitions remain far off resonance because of the large rotational level spacing characterized by the rotational constant $B_v$.  As a result, the microwave dressing enables the intermolecular interaction to be tuned independently of the optical transition, allowing the shielding-induced interaction to serve as a controllable source of few-photon nonlinearity in the cavity-QED dynamics.

In the strong-microwave regime, the weak optical couplings can be neglected, allowing the rotational subspace to be diagonalized independently. The bare rotational ground state is thereby dressed into $|+\rangle=\mu|0,0\rangle+\nu|1,0\rangle$, which is a coherent superposition of the opposite-parity rotational states with $J=0$ and $J=1$. The mixing coefficients,
$\mu=\sqrt{(1-\delta/\Omega_{\rm eff})/2}$ and $\nu=\sqrt{(1+\delta/\Omega_{\rm eff})/2}$, are determined by the effective Rabi frequency
$\Omega_{\rm eff}=\sqrt{\delta^2+\Omega_m^2}$, with the corresponding dressed-state energy $\hbar \delta_+=\hbar (\delta+\Omega_{\rm eff})/2$. Projecting the dipole-dipole interaction onto the upper dressed channel $|+,+\rangle$ and retaining the leading second-order virtual couplings to the remaining dressed channels yields the effective shielding potential derived in Appendix~\ref{appA}~\cite{PhysRevLett.130.183001, b8pm-3prn}
\begin{align}\label{potential}
V_{\rm eff}({\bf r})&=\frac{C_3}{r^3}(3\cos^2\theta-1) + \frac{{C_6}}{r^6} \sin^2{\theta}\cos^2{\theta} \nonumber \\
& +\frac{{C_6}}{36r^6(1+\delta_r^2)}(3\cos^2{\theta}-1)^2,
\end{align}
which consists of an anisotropic dipolar interaction together with a short-range repulsive shielding core of van der Waals character. The interaction coefficients are $C_3=-d^2/[24\pi\epsilon_0(1+\delta_r^2)]$ and $C_6=d^4/[32\pi^2\epsilon_0^2\hbar\Omega_m(1+\delta_r^2)^{3/2}]$, where $d$ is the permanent electric dipole moment, $\epsilon_0$ is the vacuum permittivity,
and $\delta_r=|\delta/\Omega_m|$. Compared with the$\sigma^{+}$-polarized microwave-shielding schemes~\cite{PhysRevX.13.031032,PhysRevLett.130.183001}, the $\pi$-polarized configuration enhances the effective two-body interaction by a factor of two in the dipolar coefficient $C_3$ and by a factor of four in the van der Waals coefficient $C_6$, thereby providing substantially stronger interaction-induced nonlinearities.

Figure~\ref{model}(d) compares the effective interaction potential $V_{\rm eff}$ obtained from numerical diagonalization with the analytical expression in Eq.~(\ref{potential}) for two molecules prepared in the highest microwave-dressed state at $\theta=\pi/2$. Excellent agreement is obtained over the entire interaction range, including the short-range regime $r\leq[d^2/(4\pi\epsilon_0\hbar\Omega_m)]^{1/3}=1662\,a_B$~\cite{PhysRevLett.130.183001}. The validity of the effective description is further confirmed by the near-unity overlap $\mathcal{O}$ between the exact eigenstate and the product state $|+,+\rangle$. These results demonstrate that Eq.~(\ref{potential}) provides an accurate description of the microwave-shielded interaction in the regime of interest. Moreover, the interaction is highly anisotropic, with both its magnitude and sign controlled by the intermolecular separation $r$ and the polar angle $\theta$, providing a versatile interaction resource for the cavity-QED dynamics investigated below.

To access the strong-interaction regime relevant for cavity-QED nonlinearities, we focus on the short-range subwavelength regime $r/\lambda\ll1$~\cite{doi:10.1126/science.adh3023}, where microwave shielding generates pronounced two-body interactions [Fig.~\ref{model}(d)]. For a representative separation $r=1600\,a_B$ (84.6 nm) at $\theta=\pi/2$, the repulsive interaction reaches $V_{\rm eff}/\kappa=25.3$ while maintaining a high state overlap $\mathcal{O}=0.99$. This interaction strength exceeds the cavity linewidth by more than one order of magnitude and therefore provides the microscopic origin of the interaction-induced spectral anharmonicity exploited below to realize interaction-enhanced photon blockade. The separation $r=1600\,a_B$ is chosen as a representative parameter within the strong-interaction regime, where the effective shielding potential remains quantitatively accurate and the interaction-induced nonlinearity is fully developed. Current optical-tweezer platforms have enabled deterministic preparation, coherent manipulation, and entanglement of individual polar molecules at programmable micrometer-scale separations~\cite{doi:10.1126/science.adf8999,Ruttley2025}. Nevertheless, stabilizing molecular pairs at the deeply subwavelength separation considered here remains experimentally challenging. Recent advances toward nanoscale confinement in dipolar atomic systems~\cite{doi:10.1126/science.adh3023} suggest possible routes toward accessing such interaction-dominated configurations, although extending these techniques to polar molecules remains an important open challenge.

The microwave-shielded cavity-coupled molecular pair is then described by the effective Hamiltonian ($\hbar=1$)
\begin{eqnarray}\label{Ising}
\hat{\cal H} &=&{\Delta}_{c}\hat{a}^{\dagger}\hat{a}+  \frac{1}{2}\sum_{j=1,2}(\Delta_m\hat{\sigma}_{j}^{z} +\Omega\hat{\sigma}_{j}^{x})  +  \frac{V_{\rm eff}}{4} \hat{\sigma}_{1}^{z}\hat{\sigma}_{2}^{z} \nonumber\\
&+&  {g}\hat{a}^{\dagger}[\hat{\sigma}^{-}_{1}+\cos(k r)\hat{\sigma}_{2}^-] + {\rm H.c.},
\end{eqnarray}
where $\hat{a}$ ($\hat{a}^{\dagger}$) is the annihilation (creation) operator of the cavity mode. The Pauli operators $\hat{\sigma}^{x,y,z}_j$ act on the effective two-level subspace $\{|g\rangle_j,|e\rangle_j\}$ of the $j$-th molecule, where $|g\rangle\equiv |+\rangle$ denotes the microwave-dressed ground state. Here $\hat{\sigma}^{-}=|g\rangle\langle e|$, $\hat{\sigma}^{+}=|e\rangle\langle g|$, and $\hat{\sigma}^{z}=|e\rangle\langle e|-|g\rangle\langle g|$. The detuning $\Delta_{c}$ is the cavity-pump detuning, and $\Delta_m=\Delta-V_{\rm eff}/2-\delta_+$ is the effective light-molecule detuning. The effective molecule-cavity coupling and pump Rabi frequency are renormalized by microwave dressing as ${g}=\mu g_0$ and $\Omega=\mu \Omega_0$, respectively. The wave vector is $k=2\pi/\lambda$, with $\lambda$ being the cavity wavelength. Equation~(\ref{Ising}) contains two distinct forms of geometry-dependent control: the shielding-induced interaction $V_{\rm eff}(\mathbf{r})$ determines the strength of the Ising-type molecular nonlinearity, whereas the position-dependent phase factor $\cos(kr)$ controls the relative coupling of the two molecules to the cavity field.

For molecules separated by an integer multiple of the cavity wavelength ($r=n\lambda$), the many-body Hamiltonian in Eq.~(\ref{Ising}) reduces to a Tavis-Cummings-Ising model when only the nearest-neighbor dipolar interaction is retained. In this long-distance regime, the anisotropic shielding contribution proportional to $C_6$ in Eq.~(\ref{potential}) becomes negligible because the intermolecular separation ($r>10^4\,a_B$) is much larger than the characteristic range of the short-range shielding core ($\lesssim2\times10^3\,a_B$). The resulting model is closely related to recently studied cavity-coupled Rydberg atom arrays exhibiting nonlocal dynamics and superradiant phase transitions~\cite{PhysRevLett.133.243401,PhysRevLett.134.213604}. In contrast, the present work focuses on the short-range shielding regime, where the $C_6$-induced interaction dominates and gives rise to qualitatively different cavity-QED nonlinearities.

To describe the dissipative dynamics of the cavity-coupled molecular system, we employ the Lindblad master equation,
\begin{equation}\label{master_equation}%
{ \frac{d\rho}{dt}}= -i [\hat{\cal H}, {\rho}] + {\kappa} \mathcal {D}[\hat{a}]\rho + {\gamma}\sum_{j=1}^2 \mathcal
{D}[\hat{\sigma}_{j}^-]\rho,
\end{equation}
where $\rho$ is the density matrix of the cavity-coupled molecular pair and $\mathcal {D}[\hat{o}]\rho=\hat{o} {\rho} \hat{o}^\dag - (\hat{o}^\dag \hat{o}{\rho} + {\rho}\hat{o}^\dag \hat{o})/2$ denotes the standard Lindblad dissipator. The nonclassical properties of the emitted cavity field are characterized by the second-order correlation function $g^{(2)}(\tau)$. Photon blockade is identified by sub-Poissonian equal-time statistics, $g^{(2)}(0)<1$, while photon antibunching additionally requires $g^{(2)}(0)<g^{(2)}(\tau)$.

For the numerical calculations, we consider $^{23}\mathrm{Na}^{133}\mathrm{Cs}$ polar molecules with permanent electric dipole moment $d=4.75$ Debye~\cite{PhysRevLett.130.113002}, confined in harmonic traps with trapping frequency $\omega=2\pi\times1$ MHz. The microwave field has Rabi frequency $\Omega_m=2\pi\times5$ MHz and relative detuning $\delta/\Omega_m=-0.1$. We take the spontaneous emission rate $\gamma=0.5\kappa$, the molecule-cavity coupling strength $g=4\kappa$, and a weak optical drive $\Omega=0.4\kappa$. The remaining control parameters are the cavity and molecular detunings $\Delta_c$ and $\Delta_m$, the intermolecular separation $r$, and the polar angle $\theta$. Unless otherwise specified, these parameters place the system in the weak-driving regime while ensuring that the shielding-induced interaction exceeds the cavity linewidth, thereby allowing the interaction-induced spectral anharmonicity and dispersive decoupling discussed below to be clearly resolved.

\section{Interaction-Induced Spectral Anharmonicity}\label{spectral}

To elucidate the microscopic origin of both the interaction-enhanced photon blockade and the protection of molecular entanglement, we first analyze the excitation spectrum of the closed system. Neglecting dissipation and setting the weak optical drive to zero ($\Omega=0$), Eq.~(\ref{Ising}) reduces to the Hamiltonian $\hat{\mathcal H}_0$. The total excitation number $\hat N=\hat a^\dagger\hat a+\hat\sigma_1^+\hat\sigma_1^-+\hat\sigma_2^+\hat\sigma_2^-$ is conserved, $[\hat N,\hat{\mathcal H}_0]=0$, allowing the Hilbert space to be decomposed into excitation manifolds with fixed excitation number. To reveal the role of the shielding interaction, we introduce the collective molecular basis consisting of the symmetric and antisymmetric single-excitation states
$|e_\pm\rangle=(|eg\rangle\pm|ge\rangle)/\sqrt2$
together with
$|gg\rangle$
and
$|ee\rangle$~\cite{PhysRevA.97.023831,PhysRevA.95.063842}.
In this representation, the shielding interaction directly modifies the collective molecular resonances, while the cavity couples to the symmetric and antisymmetric channels with effective coupling strengths
$g_\pm=g[1\pm\cos(kr)]$.
The single-excitation manifold therefore consists of one cavity-dark molecular state and two bright polariton states. As shown in Appendix~\ref{appB}, the corresponding single-photon resonance condition is
\begin{equation}
\Delta_c(2\Delta_m-V_{\rm eff}) = g_+^2+g_-^2.
\label{dcdm1}
\end{equation}
Equation~(\ref{dcdm1}) shows explicitly that the shielding interaction shifts the vacuum-Rabi doublet and provides direct control over the one-photon resonances. Under cavity-molecule resonance condition $\Delta_c=\Delta_m$, the resonance positions become
\begin{equation}
\Delta_{1,\pm1} = \frac{V_{\rm eff} \mp \sqrt{V_{\rm eff}^{2} + 8(g_+^2+g_-^2)}}{4}.
\label{eqdc}
\end{equation}

The one-photon resonance condition determines where the first photon can resonantly enter the cavity-molecule system. Photon blockade, however, requires the subsequent excitation to remain off resonance. The relevant quantity is therefore not the absolute position of the polariton branch, but the energetic separation between the one-photon resonance and the nearest transitions in the higher-excitation manifolds. This spectral separation is directly controlled by the shielding interaction. Although the complete multi-excitation spectrum is not analytically tractable, the high-symmetry limit $\cos(kr)=1$ already captures the essential mechanism, as demonstrated in Appendix~\ref{appB}.

For a repulsive shielding interaction ($V_{\rm eff}>0$), the molecular double-excitation manifold is shifted to higher energy. Along the blue-detuned polariton branch, this shift moves the dominant two-photon resonance farther away from the one-photon resonance, thereby increasing the effective spectral anharmonicity and suppressing sequential photon absorption. As a result, the system preferentially occupies the single-excitation manifold, giving rise to strong photon blockade. Microwave shielding therefore does more than simply shift the vacuum-Rabi doublet: it reshapes the spacing between successive excitation manifolds and consequently modifies the resonant excitation pathways. This interaction-controlled spectral restructuring constitutes the microscopic origin of the enhanced photon blockade and simultaneously establishes the dispersive energy landscape responsible for the Bell-state protection discussed below.

\begin{figure}
 \includegraphics[width=1\columnwidth]{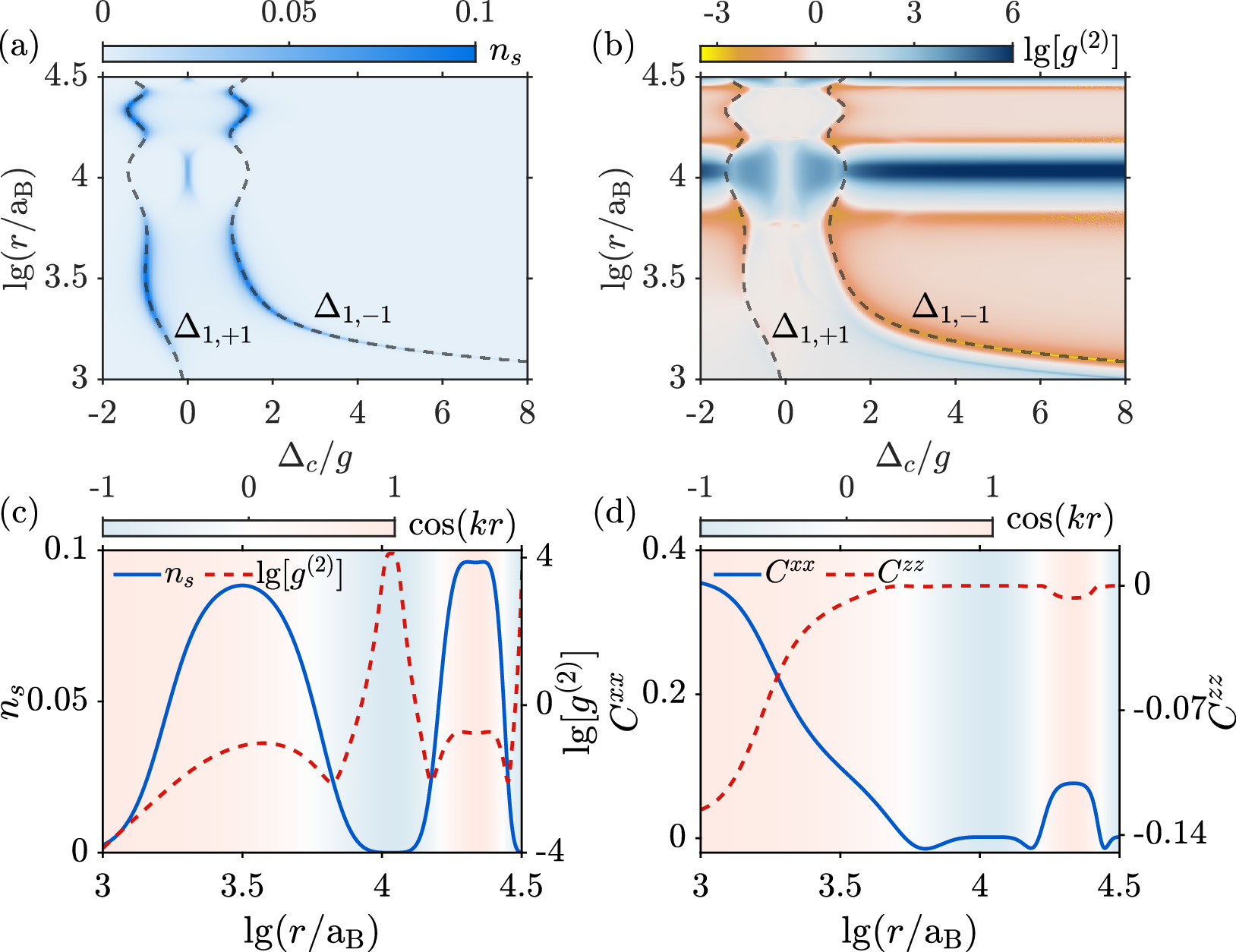}\captionsetup {labelformat=simple, labelsep=period, name=FIG.}  \caption{\label{p2} \justifying
(a) Steady-state photon number $n_{\rm s}$ and (b) second-order correlation function $g^{(2)}$ as functions of $\Delta_{\rm c}$ and $r$. The dashed curves denote the analytical single-photon resonances $\Delta_{1,\pm1}$ obtained from Eq.~(\ref{eqdc}). (c) $r$-dependence of $n_s$ and $g^{(2)}$, and (d) spin correlations $C^{\mu\mu}$ extracted along the resonance $\Delta_{1,-1}$. The pronounced negative value of $C^{zz}$ signals the interaction-induced suppression of simultaneous molecular excitations. In (c) and (d), the background shading represents $\cos(kr)$.
}
\end{figure}

\section{Interaction-Enhanced Photon Blockade}

Having established the interaction-induced spectral anharmonicity of the closed system, we now investigate its manifestation in the steady-state response of the driven-dissipative cavity-QED system. By solving the master equation, we calculate the steady-state photon number $n_{\rm s}$ and the equal-time second-order correlation function $g^{(2)}(0)$ in the $\Delta_c$-$r$ parameter plane under the resonant condition $\Delta_c=\Delta_m$. As shown in Fig.~\ref{p2}(a), the photon-number maxima closely follow the analytical single-photon resonances predicted by Eq.~(\ref{eqdc}) (dashed curves). The corresponding distribution of $g^{(2)}(0)$ in Fig.~\ref{p2}(b) exhibits pronounced antibunching along the same resonance branches, demonstrating that the steady-state photon statistics are governed by the interaction-controlled excitation spectrum. The blockade behavior is determined by the interplay of two distinct mechanisms. The optical phase factor $\cos(kr)$ controls the phase-dependent cavity coupling through the collective light-matter interaction, whereas the shielding potential $V_{\rm eff}(r,\theta)$ governs the interaction-induced spectral anharmonicity by modifying the spacing between successive excitation manifolds.

\begin{figure}
 \includegraphics[width=1\columnwidth]{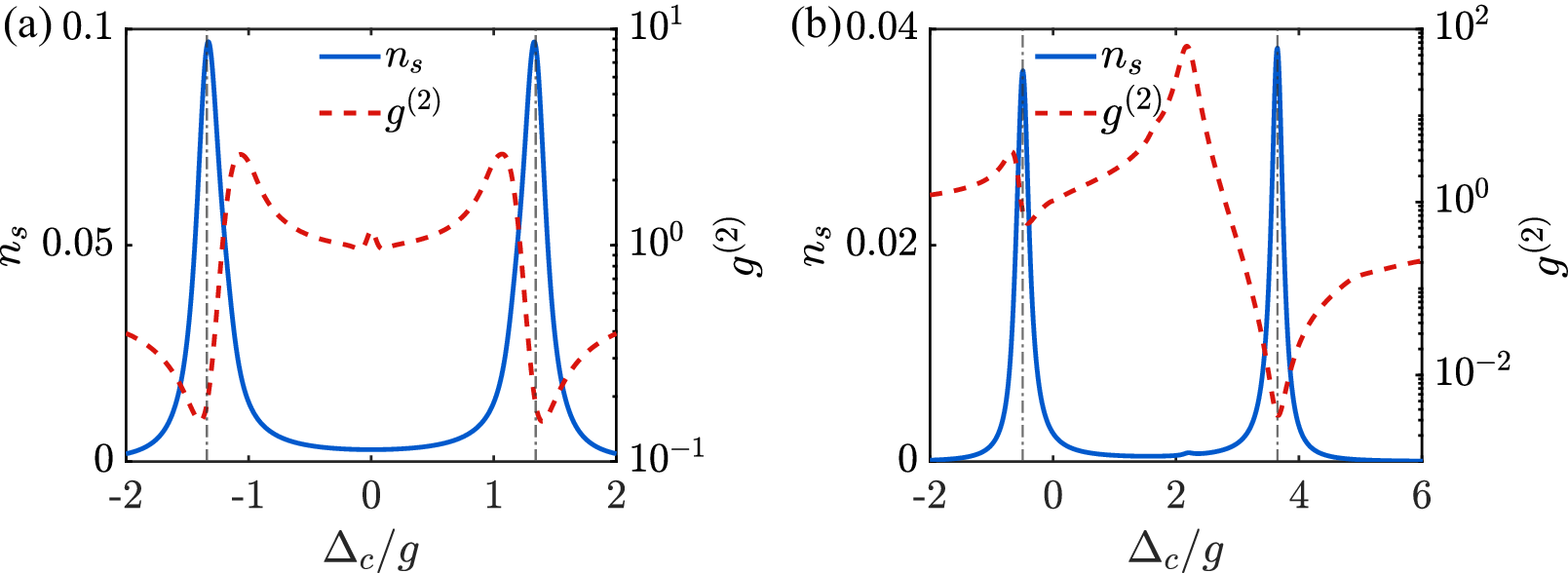}\captionsetup {labelformat=simple, labelsep=period, name=FIG.}  \caption{\label{energy_spectrum_fig3} \justifying
Steady-state photon number $n_{\rm s}$ and equal-time correlation function $g^{(2)}$ as functions of cavity detuning $\Delta_c/g$ for
(a) weak shielding, $V_{\rm eff}/\kappa=0.01$ at $r\approx2\times 10^4\,a_B$, and
(b) strong shielding, $V_{\rm eff}/\kappa=25.3$ at $r\approx1.6\times 10^3\,a_B$, with $\Delta_m=\Delta_c$.
The two separations have nearly the same phase factor, $\cos(kr)\approx0.89$, so the difference in blockade performance mainly reflects the shielding-induced interaction.
Dot-dashed lines mark the analytical single-photon resonances.
}
\end{figure}

This distinction between the phase-dependent cavity coupling and the shielding interaction is clearly illustrated in Fig.~\ref{p2}. In the long-distance regime, where the shielding interaction is negligible, the cavity response is governed almost entirely by the optical phase factor $\cos(kr)$. Near the out-of-phase condition $\cos(kr)\simeq-1$, the collective molecular mode excited by the external drive becomes nearly orthogonal to the cavity-coupled mode. As a result, the one-photon excitation channel is strongly suppressed, and the weak residual emission around $\Delta_c\simeq0$ originates predominantly from higher-order excitation processes, giving rise to pronounced photon bunching with $g^{(2)}(0)\gg1$~\cite{PhysRevA.95.063842,zhao2025tunable}. As the phase factor deviates from $\cos(kr)=-1$, the mismatch between the driven and cavity-coupled collective modes is gradually removed, leading to stronger cavity excitation and an increased steady-state photon population. In this weak-interaction regime, photon blockade is therefore governed primarily by phase-controlled interference and reaches its optimum near $\cos(kr)=0$, where destructive interference most effectively suppresses two-photon excitation pathways~\cite{PhysRevA.95.063842}.

The situation changes qualitatively in the short-range regime, where microwave shielding generates a strong effective interaction. In this regime, the shielding interaction becomes the dominant energy scale governing the excitation spectrum and strongly restructures the blue-detuned polariton branch $\Delta_{1,-1}$. As shown in Fig.~\ref{p2}(c), decreasing the intermolecular separation $r$ along this resonance branch leads to a pronounced reduction of $g^{(2)}(0)$, indicating a substantial improvement in single-photon purity. At the same time, the steady-state photon number $n_{\rm s}$ decreases because the interaction-induced shift of the molecular resonance reduces the efficiency of coherent excitation transfer to the cavity mode. Importantly, the suppression of $g^{(2)}(0)$ is considerably stronger than the accompanying reduction in $n_{\rm s}$, demonstrating that the enhanced photon blockade originates from interaction-induced spectral anharmonicity rather than from a trivial suppression of the cavity population.

To isolate the role of the shielding interaction from that of the phase-dependent cavity coupling, we compare two intermolecular separations, $r\approx1.6\times10^3\,a_B$ and $r\approx2\times10^4\,a_B$. These two configurations have nearly identical phase factors, $\cos(kr)\approx0.89$, but differ dramatically in their interaction strengths, with $V_{\rm eff}/\kappa\approx25.3$ and $0.01$, respectively. This allows the shielding interaction to be varied by more than three orders of magnitude while leaving the collective cavity coupling essentially unchanged. The corresponding photon statistics are presented in Fig.~\ref{energy_spectrum_fig3}. Despite the nearly identical phase-dependent cavity couplings, the two configurations exhibit strikingly different blockade behavior. In the weak-interaction regime ($r\approx2\times10^4\,a_B$), only moderate antibunching is obtained, with $g^{(2)}(0)\approx0.18$ near the single-photon resonance. By contrast, the strongly shielded configuration ($r\approx1.6\times10^3\,a_B$) reaches $g^{(2)}(0)\approx3\times10^{-3}$, corresponding to nearly two orders of magnitude stronger antibunching. Since the collective cavity coupling is essentially the same in both cases, this pronounced difference directly demonstrates that the enhancement of photon blockade originates from the shielding-induced interaction rather than from the phase-dependent cavity coupling.

This behavior is fully consistent with the spectral picture established in Sec.~\ref{spectral}. On the blue-detuned polariton branch, the repulsive shielding interaction enlarges the spectral anharmonicity by increasing the energy separation between the one-photon resonance and the nearest higher-excitation transitions, thereby strongly suppressing sequential photon absorption. By contrast, on the red-detuned branch the interaction can reduce this separation, facilitating multiphoton excitation and consequently weakening the blockade. The opposite behavior of the two polariton branches therefore reflects the interaction-controlled restructuring of the excitation spectrum.

\begin{figure}
 \includegraphics[width=1\columnwidth]{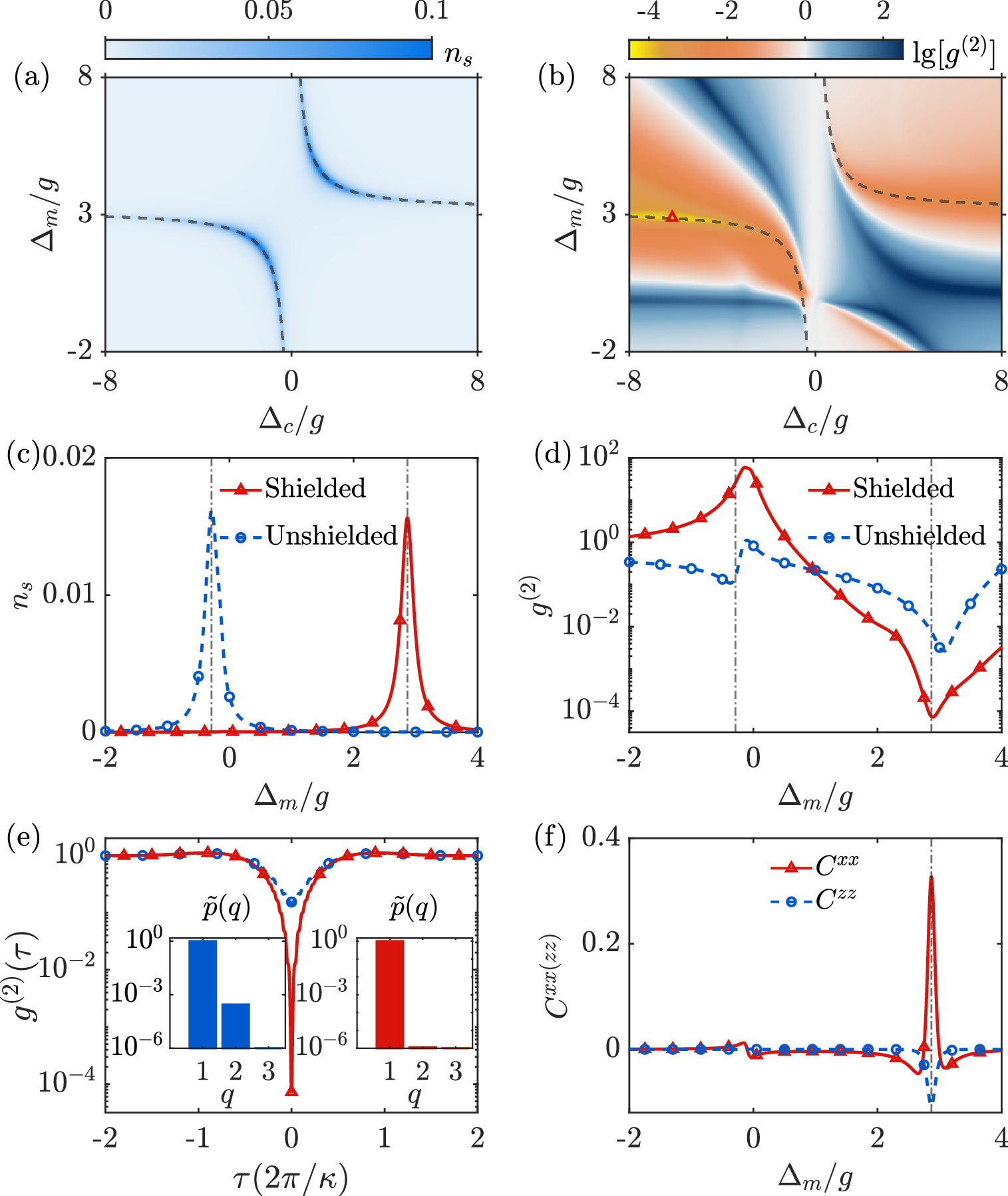}\captionsetup {labelformat=simple, labelsep=period, name=FIG.}  \caption{\label{fig3_0301}  \justifying
(a) Steady-state photon number $n_{\rm s}$ and (b) equal-time second-order correlation function $g^{(2)}$ versus $\Delta_c$ and $\Delta_m$ for $r\approx1.6\times 10^3\,a_B$. Dashed curves denote the analytical single-photon resonances from Eq.~(\ref{dcdm1}), and the red triangle marks the optimized blockade point. (c) $n_{\rm s}$ and
(d) $g^{(2)}(0)$ as functions of $\Delta_m$ at fixed $\Delta_c$. Red solid curves with triangle markers correspond to the strongly shielded configuration at $r\approx1.6\times10^3\,a_B$, while blue dashed curves with circle markers correspond to the weakly interacting configuration at $r\approx2\times10^4\,a_B$ with the same fixed $\Delta_c$. (e) Time-delayed correlation function $g^{(2)}(\tau)$ evaluated at the single-photon resonant points in (c,d). The insets show the corresponding weighted photon-number distributions $\tilde p(q)$. (f) Spin correlations $C^{xx}$ and $C^{zz}$ as functions of $\Delta_m$ for the strongly shielded configuration at $r\approx1.6\times10^3\,a_B$ and the same fixed $\Delta_c$.}
\end{figure}

Beyond modifying the photon statistics, the shielding interaction also leaves clear signatures in the microscopic intermolecular correlations. To characterize these correlations, we introduce the connected spin correlation $C^{\mu\mu}=\langle \hat{\sigma}_{1}^\mu\hat{\sigma}_{2}^\mu\rangle - \langle\hat{\sigma}_{1}^\mu\rangle \langle \hat{\sigma}_{2}^\mu\rangle$, where $\mu\in\{x,z\}$~\cite{PhysRevLett.133.106901}. Figure~\ref{p2}(d) presents the transverse ($C^{xx}$) and longitudinal ($C^{zz}$) spin correlations along the blue-detuned resonance branch $\Delta_{1,-1}$. In the weak-interaction regime, the molecules remain only weakly correlated, with $C^{xx}\approx0.07$ and $C^{zz}\approx0$ at $r\approx2\times10^4\,a_B$. As the shielding interaction increases, the transverse correlation grows markedly and reaches $C^{xx}\approx0.28$ at $r\approx1.6\times10^3\,a_B$, reflecting the enhanced coherence between the two molecular excitations. More significantly, the longitudinal correlation becomes negative, reaching $C^{zz}\approx-0.08$. A negative $C^{zz}$ indicates that simultaneous excitation of the two molecules is energetically suppressed, leading to anticorrelated molecular populations. Consequently, the occupation of the doubly excited molecular state is reduced, suppressing the dominant excitation pathway toward two-photon occupation. The evolution of the spin correlations therefore provides an independent microscopic confirmation of the interaction-enhanced blockade: microwave shielding suppresses double excitation through interaction-induced spectral anharmonicity, thereby stabilizing the single-excitation manifold.

To identify the optimal operating regime for high-purity single-photon emission, we next treat the cavity and molecular detunings, $\Delta_c$ and $\Delta_m$, as independent control parameters while fixing the intermolecular separation at the strongly shielded value $r\approx1.6\times10^3\,a_B$. Figures~\ref{fig3_0301}(a) and \ref{fig3_0301}(b) present the steady-state photon number $n_{\rm s}$ and equal-time second-order correlation function $g^{(2)}(0)$ in the two-dimensional detuning plane. Strong photon antibunching is obtained along the single-photon resonance branches, where the cavity simultaneously supports an appreciable steady-state photon population. The resonance trajectories are accurately predicted by the analytical resonance condition in Eq.~(\ref{dcdm1}), as indicated by the dashed curves, demonstrating that the steady-state response remains governed by the interaction-controlled excitation spectrum. The red triangle marks the optimal operating point, which provides the best compromise between strong antibunching and efficient single-photon emission and is used for the one-dimensional analysis below.

To directly compare the blockade performance in the strongly shielded and weakly interacting regimes, we fix the cavity detuning at the optimal operating point identified in Fig.~\ref{fig3_0301}(b) and vary the molecular detuning $\Delta_m$. The corresponding one-dimensional cuts of the steady-state photon number $n_{\rm s}$ and equal-time second-order correlation function $g^{(2)}(0)$ are presented in Figs.~\ref{fig3_0301}(c) and \ref{fig3_0301}(d). The strongly shielded configuration ($r\approx1.6\times10^3\,a_B$) is compared with the weakly interacting configuration ($r\approx2\times10^4\,a_B$) under the same cavity detuning, so that the phase-dependent cavity coupling remains nearly unchanged while the shielding-induced interaction differs by more than three orders of magnitude. The vertical dash-dotted lines indicate the analytical single-photon resonance positions predicted by Eq.~(\ref{dcdm1}).

For the strongly shielded configuration, the optimal operating point is found at $\Delta_c/g\approx-6.16$ and $\Delta_m/g\approx2.89$, where the equal-time correlation reaches $g^{(2)}(0)\approx7.7\times10^{-5}$. Under the same cavity detuning, the weakly interacting configuration exhibits only moderate antibunching, with $g^{(2)}(0)\approx0.15$. Microwave shielding therefore suppresses the equal-time correlation by more than three orders of magnitude while maintaining an appreciable steady-state photon population. This result demonstrates that the shielding-induced interaction substantially improves the purity of the emitted single photons without significantly compromising the emission brightness, identifying microwave shielding as an effective interaction resource for high-performance single-photon generation.

\begin{figure}
 \includegraphics[width=1\columnwidth]{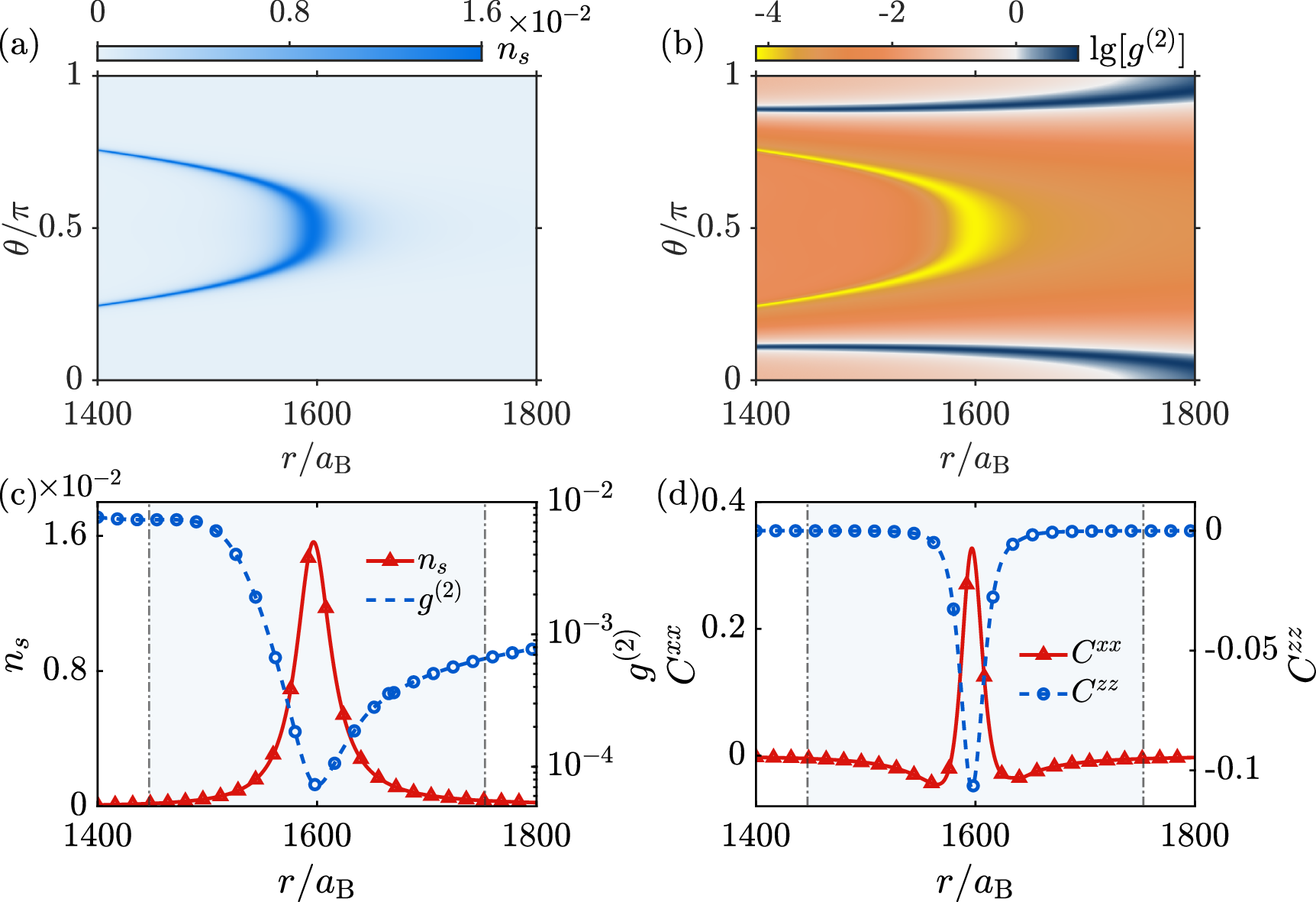}\captionsetup {labelformat=simple, labelsep=period, name=FIG.}  \caption{\label{p4} \justifying
 Spatial sensitivity of the shielding-enhanced photon blockade at $(\Delta_c/g,\Delta_m/g)=(-6.16,2.89)$.
 (a) Steady-state photon number $n_{\rm s}$ and (b) equal-time second-order correlation function $g^{(2)}$ in the $r$-$\theta$ plane. (c) $n_{\rm s}$ and $g^{(2)}$, and (d) spin correlations $C^{xx}$ and $C^{zz}$ as functions of $r$ at $\theta=\pi/2$. Shaded regions mark the relative zero-point fluctuation length, showing that the blockade is highly sensitive to molecular positional fluctuations.
}
\end{figure}

The nonclassical nature of the emitted light is further verified by the time-delayed second-order correlation function $g^{(2)}(\tau)$ shown in Fig.~\ref{fig3_0301}(e), evaluated at the optimal single-photon blockade points identified in Figs.~\ref{fig3_0301}(c) and \ref{fig3_0301}(d). Both configurations satisfy the antibunching condition $g^{(2)}(0)<g^{(2)}(\tau)$, whereas the strongly shielded configuration exhibits a substantially deeper antibunching dip, indicating a much stronger suppression of sequential photon emission. Further evidence is provided by the steady-state photon-number distribution. The insets display the weighted photon-number distribution $\tilde p(q)={q\,p(q)}/{n_{\rm s}}$, where $p(q)=\mathrm{Tr}\!\left(|q\rangle\langle q|\rho_{\rm s}\right)$. The single-photon probability increases from $\tilde p(1)\approx0.996569$ in the weakly interacting regime to $\tilde p(1)\approx0.999999$ for the strongly shielded configuration, while the residual multiphoton contribution is almost completely eliminated. The excellent agreement between the time-resolved photon correlations and the photon-number statistics  provides compelling evidence that microwave shielding realizes interaction-enhanced photon blockade through interaction-induced spectral anharmonicity, enabling near-ideal single-photon emission without severely reducing the cavity output.

Further microscopic evidence for the interaction-enhanced blockade is provided by the spin correlations shown in Fig.~\ref{fig3_0301}(f). At the optimal operating point for the strongly shielded configuration ($r\approx1.6\times10^3\,a_B$), the transverse spin correlation is significantly enhanced, reaching $C^{xx}\approx0.32$, while the longitudinal correlation becomes strongly negative, with $C^{zz}\approx-0.10$. The enhanced transverse correlation reflects the increased coherence within the single-excitation manifold, whereas the negative longitudinal correlation indicates that simultaneous excitation of the two molecules is energetically suppressed by the repulsive shielding interaction. Consequently, the population of the doubly excited molecular state is strongly reduced, blocking the dominant excitation pathway toward two-photon occupation. The spin correlations therefore provide an independent microscopic signature of the interaction-enhanced photon blockade, confirming that microwave shielding promotes coherent single-excitation dynamics while suppressing double excitation through interaction-induced spectral anharmonicity.

The interaction-enhanced photon blockade is intrinsically sensitive to the relative molecular geometry because the microwave-shielded interaction depends strongly on both the intermolecular separation $r$ and the polar angle $\theta$. To quantify this geometrical dependence, we fix the detunings at the optimal operating point $(\Delta_c/g,\Delta_m/g)\approx(-6.16,2.89)$ and calculate the steady-state photon number $n_{\rm s}$ and equal-time second-order correlation function $g^{(2)}(0)$ in the two-dimensional $r$-$\theta$ parameter space, as shown in Figs.~\ref{p4}(a) and \ref{p4}(b). Strong photon blockade is confined to a narrow region of molecular configurations, reflecting the highly anisotropic and short-range character of the shielding interaction. Consequently, the photon statistics become a sensitive probe of the relative molecular geometry.

The cuts at $\theta=\pi/2$ shown in Fig.~\ref{p4}(c) further quantify the sensitivity of the blockade to variations in the intermolecular separation. For a harmonic trap with trapping frequency $\omega=2\pi\times1~{\rm MHz}$, the relative zero-point fluctuation length is $x_{\rm ZPF}^{\rm rel}=\sqrt{\hbar/(M\omega)}=\sqrt{2}\,x_{\rm ZPF}\approx8.1~{\rm nm}\approx153\,a_B$, where $M$ is the molecular mass of ${}^{23}{\rm Na}^{133}{\rm Cs}$. A displacement comparable to this zero-point fluctuation length away from the optimal configuration substantially weakens the blockade, increasing $g^{(2)}(0)$ by several orders of magnitude while simultaneously reducing the steady-state photon number $n_{\rm s}$. These results demonstrate that the interaction-enhanced blockade is highly sensitive to molecular positional fluctuations on the nanoscale. Consequently, achieving near-ideal single-photon blockade requires tight confinement of the molecular pair around the optimal short-range configuration.

The same spatial selectivity is also reflected in the intermolecular spin correlations. As shown in Fig.~\ref{p4}(d), enhanced transverse correlations and pronounced negative longitudinal correlations appear only within the optimal blockade region, consistent with the interaction-induced suppression of simultaneous molecular excitation. This strong geometrical dependence imposes stringent requirements on molecular confinement, while at the same time suggesting that the photon statistics may provide a sensitive optical signature of nanoscale variations in the intermolecular separation. Besides the distance dependence discussed above, the polar angle $\theta$ provides an independent means of controlling the anisotropic shielding interaction. As shown in Appendix~\ref{appC}, varying $\theta$ while fixing the short-range separation at $r=1600\,a_B$ switches the optimal photon blockade between the two single-photon resonance branches. The corresponding spin correlations exhibit the same characteristic enhancement of $C^{xx}$ together with negative $C^{zz}$, confirming that suppression of double molecular excitation remains the microscopic mechanism underlying the interaction-enhanced blockade.

\begin{figure}
 \includegraphics[width=1\columnwidth]{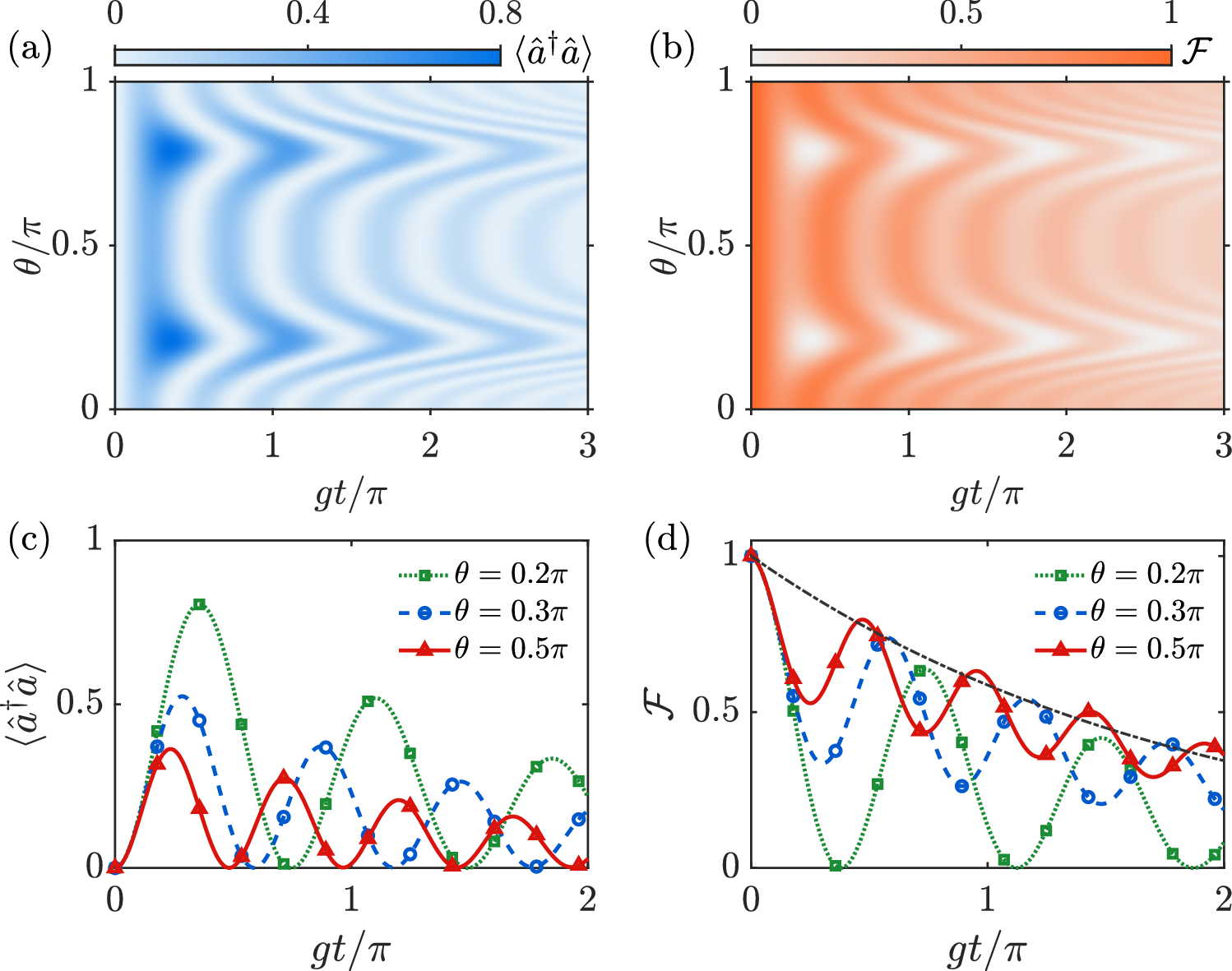}\captionsetup {labelformat=simple, labelsep=period, name=FIG.}  \caption{\label{p5} \justifying
Interaction-induced protection of molecular Bell-state fidelity at $r\approx1.6\times 10^3\,a_B$ with $\Omega=0$ and $\Delta_m=\Delta_c=0$.
 (a) Cavity photon number $\langle \hat a^\dagger \hat a\rangle$ and (b) Bell-state fidelity $\mathcal F$ as functions of time and polar angle $\theta$. (c) Time dependence of $\langle \hat a^\dagger \hat a\rangle$ and (d) $\mathcal F$ for representative angles $\theta=0.2\pi$, $0.3\pi$, and $0.5\pi$. The black dot-dashed curve in (d) shows the analytical envelope $\mathcal F_{\rm ana}$ for $\theta=0.5\pi$.
}
\end{figure}

\section{Interaction-Induced Bell-State Protection}

The shielding interaction not only enhances photon blockade but also qualitatively modifies the dissipative dynamics of molecular entanglement by restructuring the molecule-cavity resonance condition. To reveal this mechanism, we consider the undriven evolution ($\Omega=0$) at a fixed short-range separation $r\approx1.6\times10^3\,a_B$, with $\Delta_m=\Delta_c=0$. The system is initially prepared in the Bell state $|\psi(0)\rangle=(|eg\rangle+|ge\rangle)\otimes|0\rangle/\sqrt{2}$, and the preservation of the entangled state is quantified by the fidelity $\mathcal F(t)=\langle\psi(0)|\rho(t)|\psi(0)\rangle$. Figures~\ref{p5}(a) and \ref{p5}(b) present the cavity population and Bell-state fidelity in the $\theta$-$t$ plane. In the weak-interaction regime, the symmetric molecular excitation $|e_+,0\rangle$ remains nearly resonant with the cavity state $|gg,1\rangle$, allowing efficient coherent excitation exchange with the lossy cavity mode. The cavity therefore becomes transiently populated, leading to rapid cavity-mediated decoherence and a fast decay of the Bell-state fidelity. As the shielding interaction increases, the molecular excitation is shifted away from the cavity resonance, driving the system into a dispersive regime where excitation exchange with the cavity is strongly suppressed. Consequently, cavity-mediated dissipation is substantially reduced and the Bell-state lifetime is markedly prolonged.

The corresponding time evolution for three representative polar angles is presented in Figs.~\ref{p5}(c) and \ref{p5}(d). As the shielding interaction increases, the cavity population is progressively suppressed while the Bell-state fidelity is preserved over increasingly longer times. For $\theta=0.2\pi$, where $V_{\rm eff}/\kappa\approx0$, the near-resonant molecule-cavity coupling gives rise to pronounced damped oscillations in both the cavity population and the Bell-state fidelity. Increasing the shielding interaction to $V_{\rm eff}/\kappa\approx16.7$ ($\theta=0.3\pi$) and $25.3$ ($\theta=0.5\pi$) strongly suppresses the cavity occupation and substantially slows the fidelity decay.

The physical origin of this behavior can be understood from the effective single-excitation subspace $\{|e_+,0\rangle,|gg,1\rangle\}$. Since $g_-\ll g_+$ at the present intermolecular separation, the dynamics is governed almost entirely by the symmetric excitation channel. The shielding interaction introduces an effective detuning $|\Delta_{\rm eff}| \simeq {|V_{\rm eff}|}/{2}$, thereby suppressing excitation exchange between the molecular excitation and the cavity mode. In the large-detuning limit $|\Delta_{\rm eff}|\gg\{g_+,\kappa\}$, adiabatic elimination of the off-resonant cavity state then yields the effective cavity-induced decay rate $\Gamma_{\rm eff} \simeq {2\kappa g_+^2}/{V_{\rm eff}^2}$, leading to the long-time analytical fidelity
\begin{equation}
\mathcal F_{\rm ana}(t)
=
\exp[-(\gamma+\Gamma_{\rm eff})t].
\end{equation}
The analytical expression provides a good approximation to the slowly varying decay envelope of the numerically obtained fidelity for $\theta=0.5\pi$, as shown in Fig.~\ref{p5}(d). The Bell-state protection therefore originates from interaction-induced dispersive decoupling: the shielding interaction suppresses cavity-mediated dissipation by energetically decoupling the molecular excitation from the lossy cavity mode, while preserving the coherence of the molecular Bell state.

\section{Conclusion}

In conclusion, we have investigated a molecular cavity-QED platform consisting of two microwave-shielded polar molecules coupled to a single optical cavity mode. We have shown that the anisotropic shielding interaction provides a versatile and highly tunable resource for controlling both coherent and dissipative quantum dynamics. On the coherent side, the shielding interaction restructures the few-excitation spectrum and enhances its spectral anharmonicity, thereby suppressing multiphoton excitation and enabling interaction-enhanced photon blockade with near-ideal single-photon purity. The accompanying spin correlations reveal that this blockade originates microscopically from the interaction-induced suppression of simultaneous molecular excitation. Moreover, the strong dependence of the photon statistics on the intermolecular geometry highlights both the importance of tight short-range confinement and the potential of cavity photon statistics as a sensitive probe of nanoscale molecular positional fluctuations.

On the dissipative side, we have demonstrated that the same shielding interaction can effectively protect molecular entanglement by modifying the molecule-cavity resonance condition. In the strong-interaction regime, the molecular excitation is dispersively detuned from the lossy cavity mode, suppressing cavity-mediated dissipation and substantially prolonging the lifetime of an initially prepared Bell state. This establishes interaction-induced dispersive decoupling as a distinct mechanism for protecting quantum states in open molecular cavity-QED systems.

While microwave shielding has been experimentally demonstrated to suppress collisional loss and stabilize ultracold molecular gases~\cite{doi:10.1126/science.abg9502,Schindewolf2022,Bigagli2024}, our results show that it can also serve as a versatile interaction-engineering resource for generating nonclassical light and controlling cavity-mediated dissipative dynamics. More broadly, the microwave-shielded molecular cavity QED system offers a promising platform for interaction-engineered quantum nonlinear optics and dissipative many-body physics~\cite{Jia2018, Helson2023}. Extending the present scheme to larger cavity-coupled molecular arrays may further enable interaction-controlled collective quantum optical phenomena and protected many-body entangled states~\cite{Ruttley2025,kongkhambut2022observation}.

\begin{acknowledgments}
{This work was supported by the National Natural Science Foundation of China (Grant No. 12674338, Grant No.12374365, No. 12274473, No. 12135018, and No. 12574295), Guangdong Provincial Quantum Science Strategic Initiative (Grant No. GDZX2505001), National Key Research and Development Program of China (Grant No. 2021YFA0718304), and Quantum Science and Technology-National Science and Technology Major Project (Grant No.2025ZD0300400 and No. GDZX2604004), and Guangdong University of Technology SPOE Seed Foundation (SF2024111504).}
\end{acknowledgments}

\renewcommand\theequation{\Alph{section}\arabic{equation}}


\appendix

\section{Derivation of the effective shielding potential}\label{appA}
In this appendix, we obtain the adiabatic potential curves by numerical diagonalization and derive the analytical expression for the effective shielding potential. We consider a $\pi$-polarized microwave field coupling the rotational transition $|0,0\rangle  \leftrightarrow |1,0\rangle$ of each molecule. Under the rotating-wave approximation, the single-molecule Hamiltonian is
\begin{eqnarray}\label{s1}
\hat{\mathcal{H}}_1/\hbar=&&\delta(|1,1\rangle\langle1,1|+|1,0\rangle\langle1,0|+|1,-1\rangle\langle1,-1|)\nonumber\\&&+\frac{1}{2}\Omega_{\pi}(|0,0\rangle\langle1,0|+{\rm H.c.}),
\end{eqnarray} 
where $\delta=\omega_e-\omega$ is the microwave detuning, $\omega_e$ is the rotational transition frequency between the $J=0$ and $J=1$ manifolds, $\omega$ is the frequency of the $\pi$-polarized microwave field, and $\Omega_\pi$ is the corresponding Rabi frequency.
Diagonalization of $\hat{\mathcal H}_1$ leaves the two states
$|1\rangle\equiv|1,1\rangle$ and $|1'\rangle\equiv|1,-1\rangle$
uncoupled, and gives two microwave-dressed states
$|+\rangle=\mu|0,0\rangle+\nu|1,0\rangle$ and
$|-\rangle=\mu|1,0\rangle-\nu|0,0\rangle$, where $\mu=\sqrt{(1-\delta/\Omega_{\rm eff})/2}$, $\nu=\sqrt{(1+\delta/\Omega_{\rm eff})/2}$, and $\Omega_{\rm eff}=\sqrt{\delta^2+\Omega_{\pi}^2}$ is the effective Rabi frequency. The corresponding eigenenergies are $E_{1}=E_{1'}=\hbar\delta$ and $E_{\pm}=\hbar(\delta\pm\Omega_{\rm eff})/2$. Relative to the upper dressed state $|+\rangle$, the lower dressed state $|-\rangle$ has energy $-\hbar\Omega_{\rm eff}$, while the spectator states $|1\rangle$ and $|1'\rangle$ both have energy $\hbar(\delta-\Omega_{\rm eff})/2$.

The two-molecule Hamiltonian used to obtain the adiabatic potential curves is
\begin{eqnarray}\label{s2}
\hat{\mathcal{H}}_2\approx \hat{\mathcal{H}}_1\otimes I+I\otimes\hat{\mathcal{H}}_1+V({\bf r}),
\end{eqnarray}
where the relative kinetic energy is neglected under the Born-Oppenheimer approximation.
The dipole-dipole interaction takes the form
\begin{eqnarray}\label{s3}
V({\bf r})=-\frac{d^{2}}{4\pi\epsilon_{0}r^{3}}\sqrt{\frac{24\pi}{5}}\sum_{m=-2}^{2}Y_{2,m}^*(\hat{\bf r})\Sigma_{2,m},
\end{eqnarray}
where $r=|{\bf r}|$, $\hat{\bf r}={\bf r}/r$, and $Y_{2,m}^*(\hat{\bf r})$ are spherical harmonics. The operators $\Sigma_{2,m}$ form a rank-2 spherical tensor with components
\begin{eqnarray}\label{sum22}
\Sigma_{ 2,0}&=&\frac{1}{\sqrt{6}}(\hat{d}_{1}^{+}\hat{d}_{2}^{-}+\hat{d}_{1}^{-}\hat{d}_{2}^{+}+2\hat{d}_{1}^{0}\hat{d}_{2}^{0}),\nonumber\\
\Sigma_{ 2,\pm1}&=&\frac{1}{\sqrt{2}}(\hat{d}_{1}^{\pm}\hat{d}_{2}^{0}+\hat{d}_{1}^{0}\hat{d}_{2}^{\pm}),\nonumber\\
\Sigma_{ 2,\pm2}&=&\hat{d}_{1}^{\pm}\hat{d}_{2}^{\pm}.
\end{eqnarray}
Here, $\hat{d}_{j}^{\mu}$ denotes the dimensionless spherical component of the dipole operator of molecule $j$, satisfying $\hat{d}_{j}^{-}=-(\hat{d}_{j}^{+})^\dagger$.

To obtain adiabatic potential curves, we diagonalize $\hat{\mathcal{H}}_2$ in the two-molecule Hilbert space $\mathcal{H}_1\otimes \mathcal{H}_1$, where $\mathcal{H}_1$ is spanned by the four single-molecule internal states
$\{|0,0\rangle, |1,1\rangle, |1,0\rangle, |1,-1\rangle\}$. In the microwave rotating frame, the operators $\hat{d}_{j}^{\pm, 0}$ are
\begin{eqnarray}\label{s5}
\hat{d}^{0}&&=\frac{1}{\sqrt{3}}(|1,0\rangle \langle 0,0|e^{i\omega t}+{\rm H.c.}),\nonumber\\
\hat{d}^{+}&&=\frac{1}{\sqrt{3}}(|1,1\rangle \langle 0,0|e^{i\omega t}-|0,0\rangle \langle 1,-1|e^{-i\omega t}),\nonumber\\
\hat{d}^{-}&&=-(\hat{d}^{+})^\dagger.
\end{eqnarray}
Substituting Eq.~\eqref{s5} into Eq.~\eqref{sum22} and  applying the rotating-wave approximation (neglecting high-frequency oscillations), the components of the rank-$2$ tensor become
\begin{eqnarray}
\Sigma_{2,0}&&=\frac{1}{3\sqrt{6}}(2|1,0\rangle \langle 0,0|\otimes |0,0\rangle \langle 1,0|-|1,1\rangle \langle 0,0|\nonumber\\&&\otimes |0,0\rangle \langle 1,1|-|0,0\rangle \langle 1,-1|\otimes |1,-1\rangle \langle 0,0|+{\rm H.c.}),\nonumber
\end{eqnarray}
\begin{eqnarray}\label{s6}
\Sigma_{2,1}&&=\frac{1}{3\sqrt{2}}(|1,1\rangle \langle 0,0|\otimes |0,0\rangle \langle 1,0|-|0,0\rangle \langle 1,-1|\nonumber\\&&\otimes |1,0\rangle \langle 0,0|+|0,0\rangle \langle 1,0|\otimes |1,1\rangle \langle 0,0|-|1,0\rangle \langle 0,0|\nonumber\\&&\otimes |0,0\rangle \langle 1,-1|),\nonumber\\
\Sigma_{2,-1}&&=-\Sigma_{2,1}^\dagger,\nonumber\\
\Sigma_{2,2}&&=-\frac{1}{3}(|1,1\rangle \langle 0,0|\otimes |0,0\rangle \langle 1,-1|+|0,0\rangle \langle 1,-1|\nonumber\\&&\otimes |1,1\rangle \langle 0,0|),\nonumber\\
\Sigma_{2,-2}&&=\Sigma_{2,2}^\dagger.
\end{eqnarray}

When diagonalizing $\hat{\mathcal H}_2$, it is convenient to work in the dressed-state basis
$\{|+\rangle,|-\rangle,|1\rangle,|1'\rangle\}$,
where $\hat{\mathcal H}_1$ is diagonal. The effective shielding potential is obtained by tracking the adiabatic branch that correlates, in the large-distance limit, with the entrance channel in which both molecules occupy the upper dressed state, namely $|+,+\rangle$. This state is symmetric under exchange of the two molecules. Since $\hat{\mathcal H}_2$ is invariant under molecular exchange, the two-molecule Hilbert space can be decomposed into symmetric and antisymmetric sectors, and the branch connected to $|+,+\rangle$ can be obtained entirely within the symmetric sector.

Within the symmetric sector, the seven-dimensional subspace $\mathcal S_7$ containing $|+,+\rangle$ is decoupled from the remaining three-dimensional symmetric subspace. We choose the basis of $\mathcal S_7$ as
$|+,+\rangle$, $|+,1\rangle_{\rm s}$, $|+,1'\rangle_{\rm s}$,
$|+,-\rangle_{\rm s}$, $|-,1\rangle_{\rm s}$,
$|-,1'\rangle_{\rm s}$, and $|-,-\rangle$, where
$|i,j\rangle_{\rm s}=(|i,j\rangle+|j,i\rangle)/\sqrt{2}$ for $i\neq j$. In this ordered basis, the energy offsets measured from the asymptotic state
$|+,+\rangle$ are summarized by
\begin{equation}
\mathcal E
=
\mathrm{diag}(\varepsilon_1,\ldots,\varepsilon_7)
=
\mathrm{diag}(0,\Lambda,\Lambda,-\Omega_{\rm eff},
\Xi,\Xi,-2\Omega_{\rm eff}),\nonumber
\end{equation}
where $\Lambda=\frac{1}{2}(\delta-\Omega_{\rm eff})$ and $\Xi=\frac{1}{2}(\delta-3\Omega_{\rm eff})$.
Thus the diagonal contribution in $\mathcal S_7$ is $\hbar\mathcal E$.

Projecting the rank-$2$ tensor components $\Sigma_{2,m}$ onto the same ordered basis, we obtain the following independent matrix representations:
\begin{widetext}
\begin{align}
\Sigma_{2,0}^{(\mathcal S_7)}=\frac{1}{3\sqrt{6}}\left(\begin{array}{ccccccc}
4\mu^{2}\nu^{2} & 0 & 0 & 2\sqrt{2}\mu\nu w & 0 & 0 & -4\mu^{2}\nu^{2}\\
0 & -\mu^{2} & 0 & 0 & \mu\nu & 0 & 0\\
0 & 0 & -\mu^{2} & 0 & 0 & \mu\nu & 0\\
2\sqrt{2}\mu\nu w & 0 & 0 & 2w^{2} & 0 & 0 & -2\sqrt{2}\mu\nu w\\
0 & \mu\nu & 0 & 0 & -\nu^{2} & 0 & 0\\
0 & 0 & \mu\nu & 0 & 0 & -\nu^{2} & 0\\
-4\mu^{2}\nu^{2} & 0 & 0 & -2\sqrt{2}\mu\nu w & 0 & 0 & 4\mu^{2}\nu^{2}
\end{array}\right),\nonumber
\end{align}
\end{widetext}

\begin{widetext}
\begin{align}
\Sigma_{2,1}^{(\mathcal S_7)}=\frac{1}{3\sqrt{2}}\left(\begin{array}{ccccccc}
0 & 0 & -\sqrt{2}\mu^{2}\nu & 0 & 0 & \sqrt{2}\mu\nu^{2} & 0\\
\sqrt{2}\mu^{2}\nu & 0 & 0 & \mu w & 0 & 0 & -\sqrt{2}\mu^{2}\nu\\
0 & 0 & 0 & 0 & 0 & 0 & 0\\
0 & 0 & -\mu w & 0 & 0 & \nu w & 0\\
-\sqrt{2}\mu\nu^{2} & 0 & 0 & -\nu w & 0 & 0 & \sqrt{2}\mu\nu^{2}\\
0 & 0 & 0 & 0 & 0 & 0 & 0\\
0 & 0 & \sqrt{2}\mu^{2}\nu & 0 & 0 & -\sqrt{2}\mu\nu^{2} & 0
\end{array}\right),\nonumber
\end{align}
\end{widetext}

\begin{widetext}
\begin{align}
\Sigma_{2,2}^{(\mathcal S_7)}=-\frac{1}{3}\left(\begin{array}{ccccccc}
0 & 0 & 0 & 0 & 0 & 0 & 0\\
0 & 0 & \mu^{2} & 0 & 0 & -\mu\nu & 0\\
0 & 0 & 0 & 0 & 0 & 0 & 0\\
0 & 0 & 0 & 0 & 0 & 0 & 0\\
0 & 0 & -\mu\nu & 0 & 0 & \nu^{2} & 0\\
0 & 0 & 0 & 0 & 0 & 0 & 0\\
0 & 0 & 0 & 0 & 0 & 0 & 0
\end{array}\right).\nonumber
\end{align}
\end{widetext}
Here $w=\mu^{2}-\nu^{2}$. The remaining components are obtained from
$\Sigma_{2,-1}^{(\mathcal S_7)}=-(\Sigma_{2,1}^{(\mathcal S_7)})^\dagger$ and
$\Sigma_{2,-2}^{(\mathcal S_7)}=(\Sigma_{2,2}^{(\mathcal S_7)})^\dagger$. Therefore, in the basis of $\mathcal S_7$, the projected two-molecule Hamiltonian is
\begin{eqnarray}\label{s9}
\hat{\mathcal{H}}_2^{(\mathcal S_7)}
=
\hbar\mathcal{E}
-
\frac{d^{2}}{4\pi\epsilon_{0}r^{3}}
\sqrt{\frac{24\pi}{5}}
\sum_{m=-2}^{2}
Y_{2m}^{*}(\hat{\bf r})\,
\Sigma_{2,m}^{(\mathcal S_7)} .
\end{eqnarray}
Numerical diagonalization of Eq.~\eqref{s9} yields the adiabatic potential curves. The effective microwave-shielding potential is defined by the adiabatic branch that asymptotically connects to $|+,+\rangle$ as $r\rightarrow\infty$.

Following Ref.~\cite{b8pm-3prn}, we derive a near-resonant analytical
expression for the effective potential by treating the dipole-dipole interaction perturbatively up to second order, and retaining the dominant
virtual processes. The first-order contribution is
\begin{eqnarray}
\langle+,+|V|+,+\rangle
=
- \frac{d^2 \Omega_\pi^2 (3\cos^2{\theta}-1)}
{24 \pi \epsilon_0 r^3 \Omega^2_{\rm eff}}.\nonumber
\end{eqnarray}
The virtual couplings from $|+,+\rangle$ to the two spectator-state channels are
\begin{eqnarray}
\langle+,+|V|+,1\rangle_{\rm s}
&=&
\frac{d^2 \sin{\theta}\cos{\theta}}
{8 \sqrt{2}\pi  \epsilon_0 r^3}
\left(1-\frac{\delta}{\Omega_{\rm eff}}\right)
\sqrt{1+\frac{\delta}{\Omega_{\rm eff}}},
\nonumber\\
\langle+,+|V|+,1'\rangle_{\rm s}
&=&
-\frac{d^2 \sin{\theta}\cos{\theta}}
{8 \sqrt{2}\pi  \epsilon_0 r^3}
\left(1-\frac{\delta}{\Omega_{\rm eff}}\right)
\sqrt{1+\frac{\delta}{\Omega_{\rm eff}}}.\nonumber
\end{eqnarray}
These two channels give the second-order contribution
\begin{eqnarray}
V^{(2)}_{\pi,{\rm sp}}({\bf r})
&=&
\frac{|\langle+,+|V|+,1\rangle_{\rm s}|^2}
{\hbar(\varepsilon_1-\varepsilon_2)}
+
\frac{|\langle+,+|V|+,1'\rangle_{\rm s}|^2}
{\hbar(\varepsilon_1-\varepsilon_3)}
\nonumber\\
&=&
\frac{d^4 \sin^2(2\theta)}
{128 \pi^2 \epsilon_0^2 r^6 \hbar \Omega_{\pi}(1+\delta_r^2)^{3/2}},\nonumber
\end{eqnarray}
where $\delta_r=|\delta/\Omega_\pi|$.

For $\pi$ polarization, however, the first-order interaction is attractive near $\theta=0$, while the spectator-channel contribution above vanishes in this direction. It is therefore necessary to include the additional virtual channels $|+,-\rangle_{\rm s}$ and $|-,-\rangle$, whose relevant dipole-dipole matrix elements are
\begin{eqnarray}
\langle+,+|V|+,-\rangle_{\rm s}
&=&
\frac{d^2 \delta \Omega_{\pi} (3 \cos^2{\theta}-1)}
{12 \sqrt{2} \pi \epsilon_0 r^3 \Omega^2_{\rm eff}},
\nonumber\\
\langle+,+|V|-,-\rangle
&=&
\frac{d^2  \Omega_{\pi}^2 (3 \cos^2{\theta}-1)}
{24 \pi \epsilon_0 r^3 \Omega^2_{\rm eff}} .\nonumber
\end{eqnarray}
For near-resonant dressing, the contribution from $|+,-\rangle_{\rm s}$ is suppressed by $\delta_r^2$. We therefore retain the dominant additional contribution from $|-,-\rangle$. The resulting second-order correction is
\begin{widetext}
\begin{eqnarray}
V^{(2)}_{\pi}({\bf r})
&=&
\frac{|\langle+,+|V|+,1\rangle_{\rm s}|^2}
{\hbar(\varepsilon_1-\varepsilon_2)}
+
\frac{|\langle+,+|V|+,1'\rangle_{\rm s}|^2}
{\hbar(\varepsilon_1-\varepsilon_3)}
+
\frac{|\langle+,+|V|-,-\rangle|^2}
{\hbar(\varepsilon_1-\varepsilon_7)}
\nonumber\\
&=&
\frac{d^4 \sin^2{\theta}\cos^2{\theta}}
{32 \pi^2 \epsilon_0^2 r^6 \hbar \Omega_{\pi}(1+\delta_r^2)^{3/2}}
+
\frac{d^4 (3\cos^2{\theta}-1)^2}
{1152 \pi^2 \epsilon_0^2 r^6 \hbar\Omega_{\pi} (1+\delta_r^2)^{5/2}} .\nonumber
\end{eqnarray}
\end{widetext}

Combining the first- and second-order contributions, the effective potential of the microwave-shielding branch for the $\pi$-polarized microwave field is
\begin{eqnarray}\label{veffpi}
V_{\rm eff}^{\pi}({\bf r})
&=&
-\frac{d^2(3\cos^2\theta-1)}
{24\pi\epsilon_0 r^3(1+\delta_r^2)}
+
\frac{d^4\sin^2\theta\cos^2\theta}
{32\pi^2\epsilon_0^2 r^6\hbar\Omega_{\pi}(1+\delta_r^2)^{3/2}}
\nonumber\\
&&+
\frac{d^4(3\cos^2\theta-1)^2}
{1152\pi^2\epsilon_0^2 r^6\hbar\Omega_{\pi}(1+\delta_r^2)^{5/2}} .
\end{eqnarray}

For comparison, we also summarize the corresponding result for a
$\sigma^{+}$-polarized microwave field, which couples
$|0,0\rangle\leftrightarrow |1,1\rangle$. Following the same procedure, with $\Omega_{\pi}$ replaced by $\Omega_{\sigma}$, 
one obtains the adiabatic potential curves by diagonalizing the corresponding two-molecule Hamiltonian. The effective microwave-shielding potential is similarly defined by the adiabatic branch that asymptotically connects to $|+,+\rangle$
as $r\rightarrow\infty$.

The analytical effective potential is obtained by treating the dipole-dipole
interaction perturbatively up to second order. The first-order contribution is
\begin{eqnarray}
\langle+,+|V|+,+\rangle
=
\frac{d^2 \Omega_{\sigma}^2 (3\cos^2\theta-1)}
{48\pi \epsilon_0 r^3 \Omega_{\rm eff}^2 } .\nonumber
\end{eqnarray}
The relevant virtual couplings are to the two spectator-state channels
$|+,0\rangle_{\rm s}$ and $|+,1'\rangle_{\rm s}$, where
$|0\rangle\equiv |1,0\rangle$ and $|1'\rangle\equiv |1,-1\rangle$:
\begin{eqnarray}
\langle+,+|V|+,0\rangle_{\rm s}
&=&
\frac{d^2 \sin{\theta}\cos{\theta}}
{8 \sqrt{2}\pi  \epsilon_0 r^3}
\left(1-\frac{\delta}{\Omega_{\rm eff}}\right)
\sqrt{1+\frac{\delta}{\Omega_{\rm eff}}},
\nonumber\\
\langle+,+|V|+,1'\rangle_{\rm s}
&=&
\frac{d^2 \sin^2{\theta}}
{16 \pi  \epsilon_0 r^3}
\left(1-\frac{\delta}{\Omega_{\rm eff}}\right)
\sqrt{1+\frac{\delta}{\Omega_{\rm eff}}}.\nonumber
\end{eqnarray}
Thus the second-order correction is
\begin{eqnarray}
V^{(2)}_{\sigma}({\bf r})
&=&
\frac{|\langle+,+|V|+,0\rangle_{\rm s}|^2}
{\hbar(\varepsilon_1-\varepsilon_2)}
+
\frac{|\langle+,+|V|+,1'\rangle_{\rm s}|^2}
{\hbar(\varepsilon_1-\varepsilon_3)}
\nonumber\\
&=&
\frac{d^4\sin^2 \theta(1+\cos^2 \theta)}
{128\pi^2\epsilon_0 ^2 r^6 \hbar\Omega_{\sigma}
(1+\delta_r^2)^{3/2}},\nonumber
\end{eqnarray}
where $\delta_r=|\delta/\Omega_{\sigma}|$. Combining the first- and
second-order contributions gives
\begin{eqnarray}\label{veffsigma}
V_{\rm eff}^{\sigma}({\bf r})
&=&
\frac{d^2 (3\cos^2\theta-1)}
{48\pi \epsilon_0 r^3 (1+\delta_r^2)}+
\frac{d^4 \sin^2\theta(1+\cos^2\theta)}
{128\pi^2 \epsilon_0^2 r^6 \hbar\Omega_{\sigma}
(1+\delta_r^2)^{3/2}} .\nonumber\\
\end{eqnarray}

\begin{figure}
 \includegraphics[width=1.0\columnwidth]{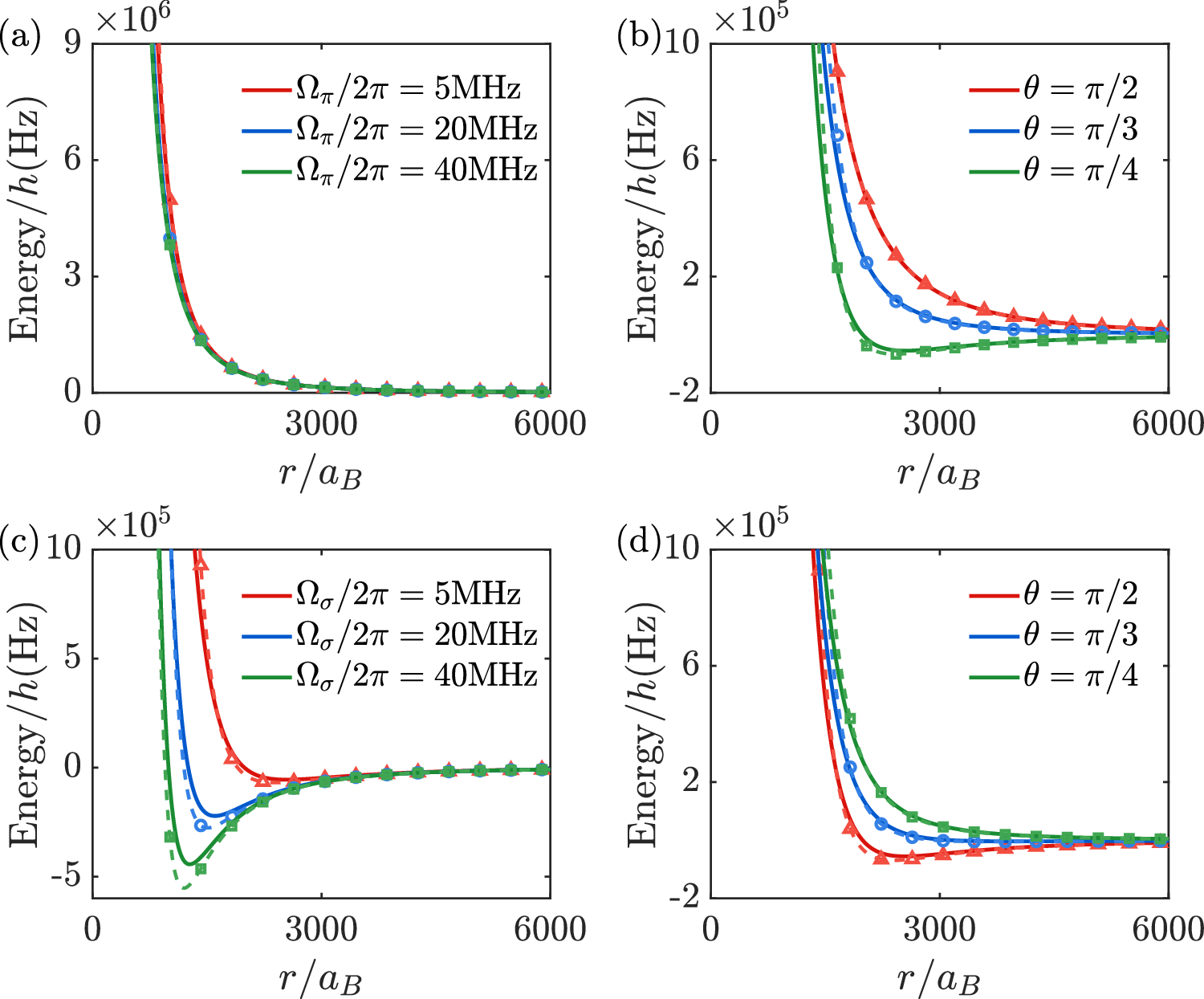}
 \captionsetup{labelformat=simple, labelsep=period, name=FIG.}
 \caption{\label{sup_potential}\justifying
 Comparison of adiabatic microwave shielding potentials obtained from
 numerical diagonalization and analytical effective potentials.
 Dark solid curves denote numerical results, while lighter curves with markers
 of the same color denote analytical results.
 (a,b) Results for the $\pi$-polarized microwave field, compared with
 Eq.~\eqref{veffpi}. 
 (c,d) Results for the $\sigma^+$-polarized microwave field, compared with
 Eq.~\eqref{veffsigma}. 
 In (a,c), the Rabi frequency is varied as
 $\Omega_{\pi,\sigma}/{2\pi}=5,20,40$ MHz at fixed $\theta=\pi/2$.
 In (b,d), the angle is varied as
 $\theta=\pi/2,\pi/3,\pi/4$ at fixed $\Omega_{\pi,\sigma}/{2\pi}=5$ MHz.
 The detuning ratio is $\delta_r=0.1$ in all panels.
 }
\end{figure}

Figure~\ref{sup_potential} compares the adiabatic shielding potentials obtained from numerical diagonalization with the analytical effective potentials derived above. In all panels, the dark solid curves denote the numerical results, while the lighter curves with markers of the same color denote the analytical expressions. Figures~\ref{sup_potential}(a) and (b) show the results for the $\pi$-polarized microwave field using Eq.~\eqref{veffpi}. In Fig.~\ref{sup_potential}(a), the Rabi frequency is
varied as $\Omega_{\pi}/{2\pi}=5,20,40$ MHz at fixed $\theta=\pi/2$ and
$\delta_r=0.1$. In Fig.~\ref{sup_potential}(b), the angle is varied as
$\theta=\pi/2,\pi/3,\pi/4$ at fixed $\Omega_{\pi}/{2\pi}=5$ MHz and $\delta_r=0.1$. The analytical effective potential shows good agreement with the numerically obtained shielding branch over the plotted range.

Figures~\ref{sup_potential}(c) and (d) present the corresponding comparison for the $\sigma^+$-polarized microwave field using Eq.~\eqref{veffsigma}. Figure~\ref{sup_potential}(c) shows the dependence on $\Omega_{\sigma}$, while Fig.~\ref{sup_potential}(d) shows the angular
dependence. The comparison confirms that the perturbative effective-potential description captures the adiabatic shielding potentials for both polarizations. This validates the effective interaction used in the cavity-QED model in the main text, where the $\pi$-polarized shielding potential provides the microscopic source of both interaction-induced spectral anharmonicity and dispersive decoupling.

\begin{figure}
 \includegraphics[width=1\columnwidth]{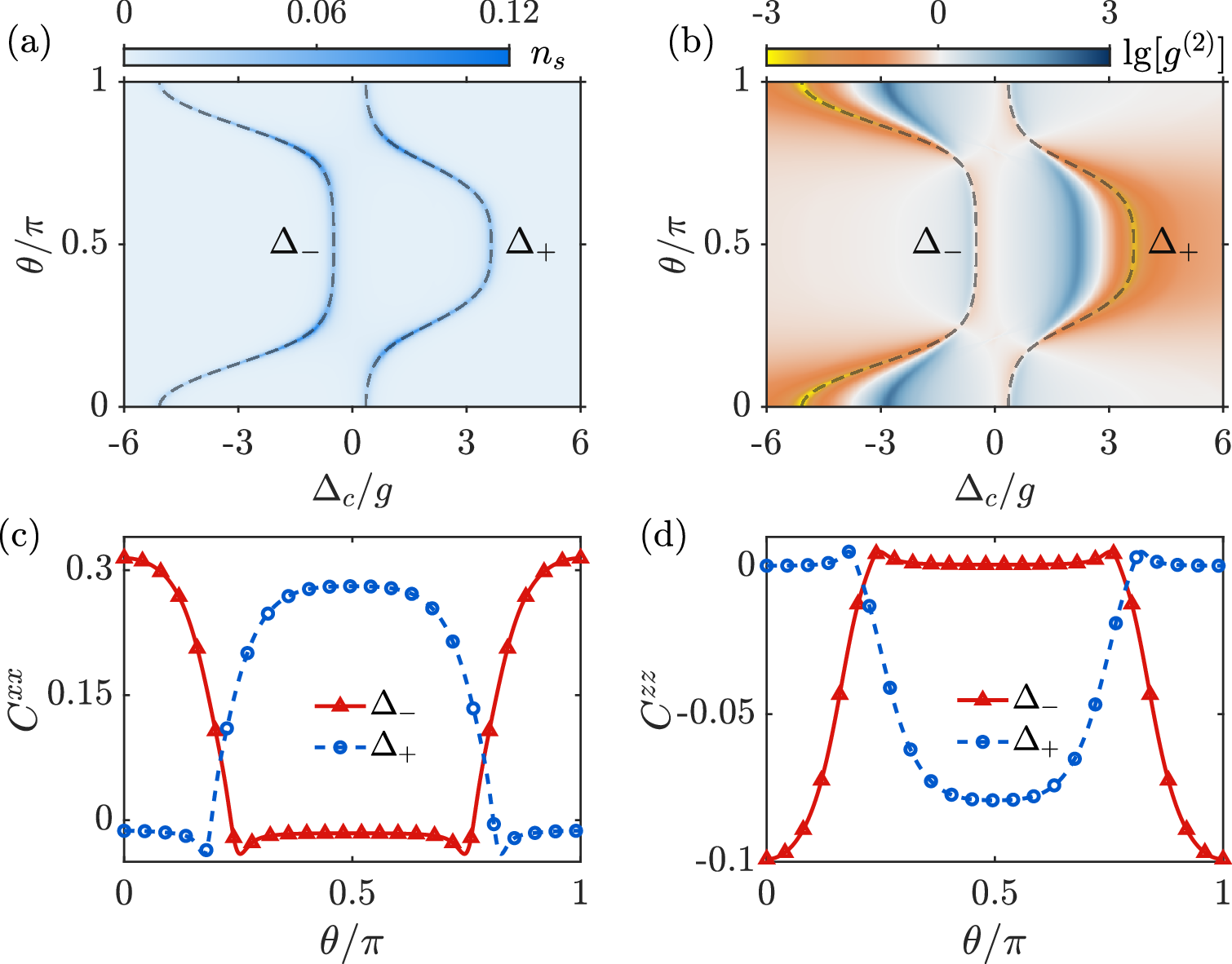}\captionsetup {labelformat=simple, labelsep=period, name=FIG.}  \caption{\label{fig5_theta_dc_ns_g20_cxx_czz} \justifying
 Angular control of shielding-induced photon blockade at fixed distance
 $r\approx1.6\times10^3\,a_B$ with $\Delta_m=\Delta_c$.
 (a) Steady-state photon number $n_{\rm s}$ and
 (b) equal-time correlation function $g^{(2)}$ in the $\Delta_c$-$\theta$
 plane. Dashed curves denote the analytical single-photon resonance branches
 $\Delta_{\pm}$.
 (c,d) Spin correlations $C^{xx}$ and $C^{zz}$ evaluated along
 $\Delta_{-}$ and $\Delta_{+}$.
}
\end{figure}

\section{Analytical Eigensystem of the Effective Hamiltonian}\label{appB}

In the weak-driving regime, we set the external pump to zero and neglect dissipation. We denote the resulting undriven Hamiltonian in Eq.~\eqref{Ising} by $\hat{\mathcal H}_0$. The total excitation number $\hat{N}=\hat{a}^{\dagger}\hat{a}+\hat{\sigma}_1^{+}\hat{\sigma}_1^{-}+\hat{\sigma}_2^{+}\hat{\sigma}_2^{-}$ is conserved, satisfying the commutation relation $[\hat{N},\hat{\mathcal{H}}_0]=0$. This allows us to analyze the
spectrum separately in each excitation manifold. To reveal the mechanism
underlying the interaction-enhanced photon blockade, we introduce the collective molecular states
$|e_+\rangle=(|eg\rangle+|ge\rangle)/\sqrt{2}$ and
$|e_-\rangle=(|eg\rangle-|ge\rangle)/\sqrt{2}$, together with
$|gg\rangle$ and $|ee\rangle$~\cite{PhysRevA.97.023831, PhysRevA.95.063842}.
For a fixed total excitation number $n\geq2$, we choose the ordered basis
\begin{equation}
\Psi_n=
\left[
|gg,n\rangle,\,
|e_+,n-1\rangle,\,
|e_-,n-1\rangle,\,
|ee,n-2\rangle
\right]^T .\nonumber
\end{equation}
In the basis $\Psi_n$, the Hamiltonian $\hat{\mathcal H}_0$ is represented by the matrix $\mathcal M=n\Delta_c I+\mathcal M_s$, where
\begin{eqnarray}\label{matrix1}
\mathcal M_s=
\left(\begin{array}{@{}cccc@{}}
0 & \sqrt{\frac{n}{2}}g_{+} & \sqrt{\frac{n}{2}}g_{-} & 0\\
\sqrt{\frac{n}{2}}g_{+} & \tilde{\Delta}-\Delta_c & 0 & \sqrt{\frac{n-1}{2}}g_{+}\\
\sqrt{\frac{n}{2}}g_{-} & 0 & \tilde{\Delta}-\Delta_c & -\sqrt{\frac{n-1}{2}}g_{-}\\
0 & \sqrt{\frac{n-1}{2}}g_{+} & -\sqrt{\frac{n-1}{2}}g_{-} & 2(\Delta_m-\Delta_{c})
\end{array}\right),\nonumber
\end{eqnarray}
The phase-dependent cavity couplings are characterized by $g_{\pm}=g[1\pm\cos(k r)]$ for the symmetric and antisymmetric channels, respectively, and $\tilde{\Delta}=\Delta_m-V_{\rm eff}/2$.

For $n=1$, the last basis state $|ee,n-2\rangle$ is absent. Diagonalization gives one dark-state eigenvalue $E_{1,0}=\tilde{\Delta}$ and two bright polariton eigenvalues
\begin{equation}
E_{1,\pm1}
=
\frac{1}{2}
\left[
\Delta_c+\tilde{\Delta}
\pm
\sqrt{(\Delta_c-\tilde{\Delta})^2+2(g_+^2+g_-^2)}
\right].
\end{equation}
The corresponding eigenstates are
\begin{eqnarray}\label{eigenstates}
|\psi_{1,0}\rangle
&=&
\mathcal{N}_0
\left(
g_- |e_+,0\rangle-g_+ |e_-,0\rangle
\right),
\nonumber\\
|\psi_{1,\pm1}\rangle
&=&
\mathcal{N}_{\pm}
\left[
(E_{1,\pm1}-\tilde{\Delta})|gg,1\rangle
+\frac{g_+ |e_+,0\rangle+g_- |e_-,0\rangle}{\sqrt{2}}
\right],\nonumber
\end{eqnarray}
where
$\mathcal{N}_0=(g_+^2+g_-^2)^{-1/2}$ and
$\mathcal{N}_{\pm}=[(E_{1,\pm1}-\tilde{\Delta})^2
+(g_+^2+g_-^2)/2]^{-1/2}$. The state $|\psi_{1,0}\rangle$ is a dark state composed purely of molecular excitations, remaining decoupled from the cavity field.

A single-photon resonance occurs when a one-excitation eigenstate is resonant with the zero-excitation state. For the bright polariton branches, this gives
\begin{eqnarray}\label{dcdm}
\Delta_{c}(2\Delta_m-V_{\rm eff})=g_+^2+g_-^2.
\end{eqnarray}
Under the equal-detuning condition $\Delta_{c}=\Delta_m$, the resonance positions reduce to
\begin{eqnarray}\label{eqdc_apB}
\Delta_{1,0}
&=&
\frac{V_{\rm eff}}{2},
\nonumber\\
\Delta_{1,\pm1}
&=&
\frac{V_{\rm eff}\mp
\sqrt{V_{\rm eff}^2+8(g_+^2+g_-^2)}}{4}.
\end{eqnarray}
Thus, the shielding interaction shifts and reshapes the vacuum-Rabi doublet by modifying the relative position of the molecular and photonic excitations.

For higher excitation manifolds ($n\geq2$), compact analytical expressions are generally not available. Simple analytical expressions can nevertheless be obtained in the high-symmetry configuration $\cos(k r)=1$, for which $g_-=0$. Under the additional condition $\Delta_c=\Delta_m$, the eigenvalues are
\begin{eqnarray}
E_{n,-1}
&=&
n\Delta_c-\frac{V_{\rm eff}}{2},
\nonumber\\
E_{n,+1}
&=&
n\Delta_c,
\nonumber\\
E_{n,\pm2}
&=&
n\Delta_c+\frac{-V_{\rm eff}\pm\chi_n}{4},\nonumber
\end{eqnarray}
where
\begin{equation}
\chi_n=\sqrt{8(2n-1)g_+^2+V_{\rm eff}^2}.\nonumber
\end{equation}
The associated eigenstates are
\begin{eqnarray}
|\psi_{n,-1}\rangle
&=&
|e_-,n-1\rangle,
\nonumber\\
|\psi_{n,+1}\rangle
&=&
\frac{1}{\sqrt{2n-1}}
\left(
\sqrt{n-1}|gg,n\rangle-\sqrt{n}|ee,n-2\rangle
\right),
\nonumber\\
|\psi_{n,\pm2}\rangle
&=&
\bar{\mathcal N}_{\pm}
\left(
\sqrt{n}g_+ |gg,n\rangle
+\sqrt{2}\lambda_{\pm}|e_+,n-1\rangle
\right.
\nonumber\\
&&
\left.
+\sqrt{n-1}g_+|ee,n-2\rangle
\right),\nonumber
\end{eqnarray}
where $\bar{\mathcal N}_{\pm}$ are normalization factors and $\lambda_{\pm}=(-V_{\rm eff}\pm\chi_n)/4$. The state $|\psi_{n,-1}\rangle$ is an antisymmetric dark state and remains decoupled from
the symmetric cavity mode.

The corresponding $n$-photon resonance positions in this high-symmetry limit are
\begin{eqnarray}
\Delta_{n,-1}
&=&
\frac{V_{\rm eff}}{2n},
\nonumber\\
\Delta_{n,+1}
&=&
0,
\nonumber\\
\Delta_{n,\pm2}
&=&
\frac{V_{\rm eff}\mp\chi_n}{4n}.
\end{eqnarray}
These expressions show explicitly how the shielding interaction changes the relative alignment between the one- and multi-photon resonances. In the noninteracting limit $V_{\rm eff}=0$, the positive- and negative-detuning resonance branches are symmetric. A finite shielding interaction breaks this symmetry. For a repulsive interaction ($V_{\rm eff}>0$), the positive-detuning single-photon branch is shifted away from the relevant two-photon resonances, which increases the spectral anharmonicity and suppresses sequential two-photon excitation. This is the regime responsible for the interaction-enhanced photon blockade discussed in the main text. On the negative-detuning side, the same interaction can reduce the spacing between the one- and two-photon resonances, thereby weakening the blockade and favoring multiphoton excitation. For an attractive interaction ($V_{\rm eff}<0$), the detuning asymmetry is reversed: the enhanced blockade is transferred to the negative-detuning side. Therefore, the role of $V_{\rm eff}$ is not merely to shift the resonance frequencies, but to restructure the relative spacing of successive excitation manifolds, which provides the microscopic origin of the interaction-enhanced photon blockade.

\section{Polar-Angle Dependence of Interaction-Enhanced Photon Blockade}
\label{appC}
The anisotropic shielding potential $V_{\rm eff}(r,\theta)$ enables
independent angular control of photon blockade. Here we fix the short-range
separation at $r\approx1.6\times10^3\,a_B$ and impose the equal-detuning condition $\Delta_m=\Delta_c$.
Figures~\ref{fig5_theta_dc_ns_g20_cxx_czz}(a) and
\ref{fig5_theta_dc_ns_g20_cxx_czz}(b) show the steady-state photon number
$n_{\rm s}$ and equal-time correlation function $g^{(2)}$ in the
$\Delta_c$-$\theta$ plane. The dashed curves denote the analytical
single-photon resonance branches $\Delta_{\pm}$ derived in
Appendix~\ref{appB}. As $\theta$ is varied, the shielding interaction changes both its magnitude and sign, thereby shifting the strong-blockade region between the two single-photon resonance branches. In the repulsive regime, strong blockade appears on one resonance branch, whereas in the attractive regime the antibunched region switches to the other branch.

The corresponding spin correlations along the two resonance branches are shown in Figs.~\ref{fig5_theta_dc_ns_g20_cxx_czz}(c) and ~\ref{fig5_theta_dc_ns_g20_cxx_czz}(d). The regions with strong blockade are accompanied by enhanced transverse correlation $C^{xx}$ and negative
longitudinal correlation $C^{zz}$. This behavior is consistent with the mechanism discussed in the main text: the negative longitudinal correlation reflects the interaction-induced suppression of simultaneous molecular excitation, which weakens the dominant pathway toward two-photon occupation. The angular dependence therefore provides a complementary means of tuning photon blockade without changing the intermolecular separation.

%

\end{document}